\documentclass[a4paper,12pt]{article}

\usepackage{graphicx}
\usepackage{amsmath}
\usepackage{amssymb}
\usepackage{multirow}
\usepackage{float} 

\begin{document}

\begin{flushright}
SHEP-12-11
\end{flushright}
\vspace*{1.0truecm}

\begin{center}
{\large\bf Scope of Higgs production in association with a bottom quark pair in probing the Higgs sector of the NMSSM at the LHC}\\
\vspace*{1.0truecm}
{\large M. M. Almarashi$^*$\footnote{al\_marashi@hotmail.com} and S. Moretti$^{**}$}\footnote{s.moretti@soton.ac.uk}\\
\vspace*{0.5truecm}
$^*${\it Department of Physics, Faculty of Science, \\
 Taibah University, P.O.Box 30002, Madinah, Saudi Arabia } \\
$^{**}${\it School of Physics \& Astronomy, \\
 University of Southampton, Southampton, SO17 1BJ, UK}
\end{center}

\vspace*{1.0truecm}
\begin{center}
\begin{abstract}
\noindent 
We review the potential of the LHC to detect a very light CP-odd Higgs boson of the NMSSM, $a_1$, through its direct production
in association with a bottom-quark pair at large tan$\beta$. We also review
the LHC discovery potential of the two lightest CP-even Higgs
states, decaying into two lighter Higgs states or into the lightest CP-odd Higgs state
and the $Z$ gauge boson.

\end{abstract}
\end{center}

\section{Introduction}

The Minimal Supersymmetric Standard Model (MSSM) \cite{MSSMreview} is probably one of the most studied
Beyond the SM (BSM) scenarios. However, 
this model suffers
from two critical flaws: the $\mu$-problem \cite{Kim:1983dt} and the little hierarchy problem. 
The former flaw results from
the fact that the Superpotential has a dimensional parameter, $\mu$
(the so-called `Higgs(ino) mass parameter'), whose natural value would be 
either 0 or $m_{\rm Pl}$ (the Planck mass). However, phenomenologically, in order to achieve Electro-Weak Symmetry Breaking (EWSB), $\mu$ is required to take values of the order of the EW scale or possibly up to the TeV range.
The latter flaw emerged first from LEP, which failed to detect a light CP-even Higgs boson, $h$, thereby imposing 
severe constraints on $m_h$. 
For this kind of Higgs state to
pass the experimental constraints, large higher order corrections from both the SM and SUSY particle spectrum are required. 
The largest contributions come from the third generation, top quarks and squarks. However, these required large corrections seem
quite unnatural. Recall in fact that at tree level the lightest CP-even Higgs boson mass of the MSSM is less than $M_Z$. Even recent
LHC results, hinting at the possible existence of a SM-like Higgs state with mass of 124--126 GeV \cite{ATLAS-Higgs,CMS-Higgs}, weaken the MSSM assumption, as
such mass values are really extreme in such a SUSY realisation,
towards the very end of the allowed mass range.

The simplest SUSY realisation beyond the MSSM that can solve these two problems at once is the NMSSM (for reviews see \cite{NMSSMreviewed1,NMSSMreviewed2}). This scenario includes a Higgs singlet Superfield in addition to the two MSSM-type Higgs
doublets, giving rise to seven Higgs states: three CP-even Higgses $h_{1, 2, 3}$ ($m_{h_1} < m_{h_2} < m_{h_3}$), 
two CP-odd Higgses $a_{1, 2}$ ($m_{a_1} < m_{a_2} $) and a pair of charged Higgses $h^{\pm}$. When the scalar
component of the singlet Superfield acquires a Vacuum Expectation Value (VEV), an `effective' $\mu$-term, $\mu_{\rm eff}$, will be
automatically generated and can rather naturally have values of order of the EW to TeV scale, as required \cite{Ellis:1988er}.
In addition, in the NMSSM, the little hierarchy problem can be relieved \cite{BasteroGil:2000bw,Dermisek:2005ar}, since a SM-like scalar Higgs boson in the NMSSM context requires
less (s)quark corrections than those in the MSSM or it can have mass less than the LEP bound due to unconventional decays
over some regions of the 
NMSSM parameter space\footnote{We will give more explanations of this in Sec. 4.}. In fact, currently, the NMSSM can also explain not only the LHC excess \cite{Ellwanger:2012ke,Ellwanger:2011aa}
a possible LEP excess and is definitely 
preferred by EW global fits 
\cite{Dermisek:2005gg,Dermisek:2006wr,excess}. 

\section{The NMSSM Superpotential}
The Superpotential of the NMSSM is given by
 \begin{equation}  
 W={\bf h_u}\hat Q\hat H_u\hat U^c-{\bf h_d}\hat Q\hat H_d\hat D^c-{\bf h_e}\hat L\hat H_d\hat E^c+\lambda\hat S\hat H_u\hat H_d
+\frac{1}{3}\kappa{\hat S}^3,
\end{equation}
where ${\bf h_u}$, ${\bf h_d}$, ${\bf h_e}$, $\lambda$ and $\kappa$ are dimensionless couplings.
The term $\lambda\hat S\hat H_u\hat H_d$ has
been introduced to solve the $\mu$-problem of the MSSM Superpotential. However, the Superpotential in Eq. (1)
without the term $\frac{1}{3}\kappa{\hat S}^3$
gives rise to an extra global $U(1)$ symmetry,
the so-called Peccei-Quinn symmetry $U(1)_{PQ}$ \cite{Peccei:1977hh, Peccei:1977ur}. Once the Higgs bosons
take on VEVs, this symmetry will break spontaneously and lead to the appearance of a CP-odd scalar, 
called a Peccei-Quinn axion. In fact, this axion has not been seen experimentally. In addition, there 
are severe astrophysical and cosmological constraints on $\lambda$, that is $10^{-7}<\lambda<10^{-10}$ \cite{Hagiwara:2002fs}.
These constraints necessitate a very large value of $<S>$ in order to solve the $\mu$-problem.
So, this is not a satisfactory way to solve the latter.

One elegant way to solve the $\mu$-problem is to break the $U(1)_{PQ}$ by introducing an additional
term in the Superpotential. This is the last term in Eq. (1) and consequently the axion can be avoided. However, introducing
this new term in the Superpotential enables one to break the PQ symmetry but the Superpotential still have
a discrete $\mathbb{Z}_3$ symmetry. This discrete symmetry is spontaneously broken
when the additional complex scalar field acquires a VEV and that will lead to
the domain wall problem. That is, during the EW phase transition of the early universe, this broken symmetry causes
a dramatic change of the universe evolution and creates unobserved large anisotropies in the cosmic microwave background \cite{Zeldovich:1974uw}.

In order to solve the domain wall problem, one needs to break the $\mathbb{Z}_3$ symmetry by introducing higher order operators at
the Plank scale. However, these operators generate quadratic tadpoles for the singlet. So, one also needs to impose
a new discrete invariance, a $\mathbb{Z}_2$ symmetry, on these operators in order to get rid of the dangerous tadpole
contributions, see \cite{NMSSMreviewed1} for more details.

\section{The Higgs sector of the NMSSM}
The NMSSM Higgs sector contains two Higgs doublets and one Higgs singlet:

\begin{equation}
H_d= \left( \begin{array}{ccc}
H^0_d \\
H^-_d
\end{array} \right), \qquad
H_u =\left( \begin{array}{ccc}
H^+_u \\
H^0_u
\end{array} \right), \qquad
S.
\end{equation}
The scalar potential for the Higgs fields can be written as \cite{MNZ}:
\begin{equation}
 V_H=V_F+V_D+V_{\rm soft},
\end{equation}
where
\begin{equation}
 V_F = {\lvert \lambda S\rvert}^2(\lvert H_u\rvert^2+\lvert H_d\rvert^2 )+\lvert \lambda H_uH_d+\kappa S^2\rvert^2,
\end{equation}

\begin{equation}
 V_D =\frac{g^2_1+g^2_2}{8}\big (\lvert H_d\rvert^2-\lvert H_u\rvert^2)^2+\frac{1}{2}g^2_2\lvert H^\dag_uH_d\rvert^2,
\end{equation}

\begin{equation}
 V_{\rm soft}=m^2_{H_u}H^\dag_uH_u+m^2_{H_d}H^\dag_dH_d+m^2_SS^\dag S 
+\big (\lambda A_\lambda SH_uH_d+\frac{1}{3}\kappa A_\kappa S^3+h.c.\big ).
\end{equation}

To generate EWSB, the Higgs fields should have VEVs. In fact, if one 
assumes that the VEVs are real and positive, they can be described by 
\begin{equation}
<H_d>= \frac{1}{\sqrt{2}}\left( \begin{array}{ccc}
\upsilon_d \\
0
\end{array} \right), \qquad
<H_u>=\frac{1}{\sqrt{2}}\left( \begin{array}{ccc}
0 \\
\upsilon_u
\end{array} \right), \qquad
<S>=\frac{1}{\sqrt{2}}\upsilon_s.
\end{equation}

At the physical minimum of the scalar potential, $V_H$, the soft mass parameters of the Higgs fields are related to the VEVs through
the following relations \cite{MNZ}:
\begin{equation}
 m^2_{H_d}=\frac{g^2_1}{8}(\upsilon^2_u-\upsilon^2_d)-\frac{1}{2}\lambda^2\upsilon^2_u
+\frac{1}{2}(\sqrt{2}A_\lambda+\kappa\upsilon_s)\lambda\upsilon_s\frac{\upsilon_u}{\upsilon_d}-\frac{1}{2}\lambda^2\upsilon^2_s,
\end{equation}
\begin{equation}
 m^2_{H_u}=\frac{g^2_1}{8}(\upsilon^2_d-\upsilon^2_u)-\frac{1}{2}\lambda^2\upsilon^2_d
+\frac{1}{2}(\sqrt{2}A_\lambda+\kappa\upsilon_s)\lambda\upsilon_s\frac{\upsilon_d}{\upsilon_u}-\frac{1}{2}\lambda^2\upsilon^2_s,
\end{equation}
\begin{equation}
 m^2_{S}=-\kappa^2\upsilon^2_s-\frac{1}{2}\lambda^2\upsilon^2+\kappa\lambda\upsilon_u\upsilon_d
+\frac{1}{\sqrt{2}}\lambda A_\lambda \frac{\upsilon_u\upsilon_d}{\upsilon_s}-\frac{1}{\sqrt{2}}\kappa A_\kappa \upsilon_s.
\end{equation}

The physical Higgs states arise after the Higgs fields acquire VEVs and rotate away the Goldstone modes. As a result,
the potential can be written as
\begin{equation}
V_{H}=m^2_{h^\pm}h^+h^-+\frac{1}{2}(P_1\ \   P_2)\mathcal{M}_P\left( \begin{array}{ccc}P_1 \\ P_2 \end{array} \right)
 +\frac{1}{2}(S_1\ \   S_2\ \ S_3)\mathcal{M}_S\left( \begin{array}{ccc}S_1 \\ S_2 \\ S_3 \end{array} \right).
\end{equation}

The masses of charged Higgs fields, $h^\pm$, at tree level are
\begin{equation}
 m^2_{h^\pm}=m^2_A+M^2_W-\frac{1}{2}(\lambda \upsilon)^2,
\end{equation}
where 
\begin{equation}
 m^2_A=\sqrt{2}\frac{\mu_{\rm eff}}{\sin2\beta}\bigg(A_{\lambda}+\frac{\kappa\mu_{\rm eff}}{\lambda}\bigg).
\end{equation}

Using the minimisation conditions, one can obtain the mass matrices in the scalar and pseudoscalar sectors. First, the mass matrix
for CP-even Higgs states at tree level has the following entries \cite{MNZ}:
\begin{equation}
 \mathcal{M}_{S11}=m^2_A+\bigg (M^2_Z-\frac{1}{2}(\lambda\upsilon)^2\bigg ){\rm sin}^22\beta,
\end{equation}
\begin{equation}
 \mathcal{M}_{S12}=-\frac{1}{2}\bigg (M^2_Z-\frac{1}{2}(\lambda\upsilon)^2\bigg ){\rm sin}4\beta,
\end{equation}
\begin{equation}
\mathcal{M}_{S13}=-\frac{1}{2}\bigg (m^2_A{\rm sin}2\beta+2\frac{\kappa{\mu^2_{\rm eff}}}{\lambda}\bigg )\bigg (\frac{\lambda\upsilon}{\sqrt{2}\mu_{eff}}\bigg ){\rm cos}2\beta, 
\end{equation}
\begin{equation}
 \mathcal{M}_{S22}=M^2_Z{\rm cos}^22\beta+\frac{1}{2}(\lambda\upsilon)^2{\rm sin}^22\beta,
\end{equation}
\begin{equation}
 \mathcal{M}_{S23}=\frac{1}{2}\bigg (4{\mu^2_{\rm eff}}-m^2_A{\rm sin}^22\beta-\frac{2\kappa{\mu^2_{\rm eff}}{\rm sin}2\beta}{\lambda}\bigg )\frac{\lambda\upsilon}{\sqrt{2}\mu_{\rm eff}},
\end{equation}
\begin{equation}
\mathcal{M}_{S33}=\frac{1}{8}m^2_A{\rm sin}^22\beta\frac{\lambda^2\upsilon^2}{\mu^2_{\rm eff}}+4\frac{\kappa^2{\mu^2_{\rm eff}}}{\lambda^2},
+\frac{\kappa A_\kappa \mu_{\rm eff}}{\lambda}-\frac{1}{4}\lambda\kappa\upsilon^2{\rm sin}2\beta.
\end{equation}
Second, the mass matrix for CP-odd Higgs states at tree level has the following entries \cite{MNZ}:
\begin{equation}
\mathcal{M}_{P11}=m^2_A,
\end{equation}
\begin{equation}
 \mathcal{M}_{P12}=\frac{1}{2}\bigg (m^2_A{\rm sin}2\beta-6\frac{\kappa{\mu^2_{\rm eff}}}{\lambda}\bigg )\frac{\lambda\upsilon}{\sqrt{2}\mu_{\rm eff}},
\end{equation}
\begin{equation}
\mathcal{M}_{P22}=\frac{1}{8}\bigg (m^2_A{\rm sin}2\beta+6\frac{\kappa{\mu^2_{\rm eff}}}{\lambda}\bigg )\frac{\lambda^2\upsilon^2}{\mu^2_{\rm eff}}{\rm sin}2\beta
-3\frac{\kappa\mu_{eff}A_\kappa}{\lambda}.
\end{equation}

To a good approximation, at large tan$\beta$ and large $m_A$, the tree level neutral Higgs boson masses are given by
the following expressions \cite{MNZ}: 
\begin{equation}
 m^2_{a_1}=-\frac{3\kappa\mu_{\rm eff}A_{\kappa}}{\lambda},
\end{equation}

\begin{equation}
 m^2_{a_2}=m^2_A\bigg (1+\frac{1}{8}(\frac{\lambda^2\upsilon^2}{\mu^2_{\rm eff}}){\rm sin}^22\beta\bigg ),
\end{equation}

\begin{eqnarray}
m^2_{h_{1/2}}&&=\frac{1}{2} \Bigg\{ 
M^2_Z+\frac{\kappa\mu_{\rm eff}}{\lambda}\bigg(\frac{4\kappa\mu_{\rm eff}}{\lambda}+A_{\kappa}\bigg) \nonumber \\
&& \mp {\sqrt{\Bigg[M^2_Z-\frac{\kappa\mu_{\rm eff}}{\lambda}\bigg(\frac{4\kappa\mu_{\rm eff}}{\lambda}+A_{\kappa}\bigg)\Bigg]^2+
\frac{\lambda^2\upsilon^2}{2{\mu^2_{\rm eff}}}\Bigg[4\mu^2_{\rm eff}-m^2_A\sin^22\beta\Bigg]^2}}\Bigg\},
\end{eqnarray}

\begin{equation}
 m^2_{h_3}=m^2_A\bigg (1+\frac{1}{8}(\frac{\lambda^2\upsilon^2}{\mu^2_{\rm eff}}){\rm sin}^22\beta\bigg ).
\end{equation}

\section{LHC phenomenology of the NMSSM Higgs sector}
Because of the existence of a singlet Superfield in the NMSSM, the latter is phenomenologically richer than the MSSM. 
In fact, the NMSSM has seven Higgs states and five neutralinos compared to only five Higgs states and four neutralinos
in the MSSM. As a consequence, the search for Higgs bosons in the context of the NMSSM at present and future colliders is
a big challenge and more complicated than in the MSSM.   

It was mentioned before that the mass of the lightest CP-even Higgs boson in the MSSM, $m_h$, at tree level should be less than
$M_Z$. So, large radiative corrections, mainly from top and stop loops, are required to pass the LEP lower limit on the Higgs mass.
In fact, to achieve this we need large stop masses, which only contribute logarithmically
in the loop corrections. This large discrepancy between top and stop masses causes essentially a fine tuning problem
\cite{NMSSMreviewed1} (the aforementioned little hierarchy problem).
  
As for the NMSSM, the situation is quite different. Assuming CP-conservation in the
Higgs sector, the upper mass bound for the lightest CP-even Higgs boson at tree level is given by
\begin{equation}
 m^2_{h1}\leq M^2_Z\bigg ({\rm cos}^2(2\beta)+\frac{2\lambda^2{\rm sin}^2(2\beta)}{g^2_1+g^2_2}\bigg).
\end{equation}
The last term in this expression can lift $m_{h_1}$ up to 10 GeV higher than the corresponding mass of the MSSM. So,
smaller loop corrections are required to pass the lower bound on the SM-like Higgs mass. However, 
since the higher order corrections are similar to those in
the MSSM, the upper mass bound reaches 135 -- 140 GeV for maximal
stop mixing and tan$\beta=2$ \cite{upper,Ellwanger:1999ji}, however, this configuration is already excluded
in the MSSM by LEP data.  Finally, notice that the corrections
to the lightest CP-even Higgs boson mass are already calculated at complete 
one loop level \cite{Ellwanger:1993hn,Elliott:1993bs,Pandita:1993hx}
and also at the dominant two loop level \cite{Ellwanger:1999ji}.  

\begin{figure}[h]
 \centering\begin{tabular}{cc}

  \includegraphics[scale=0.4]{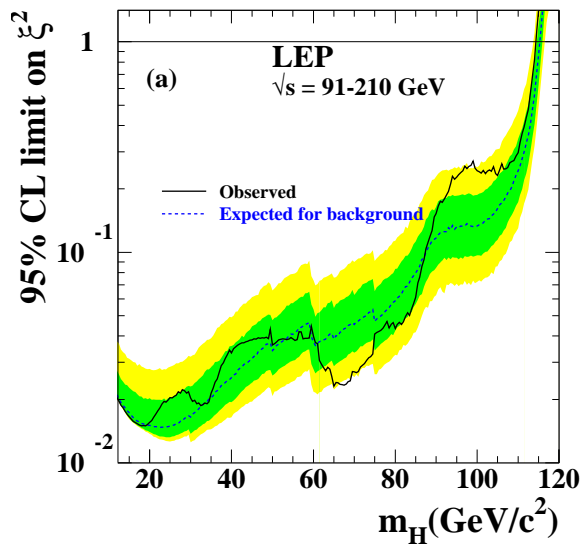}
 \end{tabular}
\caption{Upper limit on the ratio $\xi$ from LEP, where the SM Br($H\to b\bar b$) and Br($H\to \tau^+\tau^-$) are assumed.
Full line represents the observed limit and dashed line represents the expected limit. The green band
and yellow band are within 68\% and 95\% probability, respectively \cite{NMSSMreviewed2}.} 
\label{higgslep}

\end{figure}

Furthermore, the most interesting 
property of the NMSSM that can solve the little hierarchy problem of the MSSM comes from the fact that in large areas
of the NMSSM parameter space Higgs-to-Higgs decays are kinematically open. For instance, the existence of the lightest CP-odd Higgs
boson $a_1$ with mass less than $\frac{1}{2} m_{h_1}$ is quite natural in the NMSSM, see, e.g., \cite{Almarashi:2010jm}. In fact, the 
Branching ratio (Br) for the decay $h_1\to a_1a_1$, Br($h_1\to a_1a_1$), can be dominant in large regions of parameter space and as a result the Br($h_1\to b\bar b$) is suppressed. This
unconventional decay channel is so important as it could explain the $2.3 \sigma$ excess observed at LEP for a Higgs
mass, $m_{H}$, around 100 GeV as shown in figure \ref{higgslep}. The reduced coupling in the figure is defined as follows:
\begin{equation} 
 \xi^2=\bigg (\frac{g_{HZZ}}{g^{SM}_{HZZ}}\bigg )^2.
\end{equation}
Here, $g^{SM}_{HZZ}$ denotes the SM $HZZ$ coupling while $g_{HZZ}$ the non-standard coupling. As it is clear from
the plot the excess occurs when the Br$(H_{SM}\to b\bar b)$ times $\xi^2$ gives about 20\%. In the context of the NMSSM,
one can explain this excess in two ways. Firstly, a SM-like Higgs boson, $h_{1, 2}$, can decay dominantly 
into a pair of $a_1$'s and so the Br$(H\to b\bar b)$ is suppressed \cite{Dermisek:2005gg,Dermisek:2006wr,excess}. This scenario can 
relieve the little
hierarchy problem but requires that $m_{a_1} < 2 m_b$. (Notice that this mass region is currently highly constrained by 
ALEPH \cite{Schael:2010aw}
and BaBar \cite{Aubert:2009cka} data.) In fact, there is also another possibility in the NMSSM that can explain
the LEP excess due to the fact that 
the Br($a_1\to \gamma\gamma$) can be dominant when the $a_1$ is highly singlet and again, as a result, the Br($a_1\to b\bar b$) is suppressed
even with $m_{a_1}>2m_b$ \cite{Almarashi:2010jm,Almarashi:2011bf}.
Secondly, a CP-even Higgs boson, $h_1$, has a reduced coupling with $\xi\lesssim$ 0.4 \cite{Dermisek:2007ah}, due
to the mixing between the Higgs singlet and doublets. Notice that neither the SM nor the MSSM can explain such modest excess,
as they have a Br$(H/h\to b\bar b )$ which is always dominant, hence yielding an excess much above the experimental limit.
Besides, the NMSSM could also explain the recent excess observed at the LHC for a Higgs mass around 125 GeV
\cite{Ellwanger:2011aa,Gunion:2012zd,King:2012is,Vasquez:2012hn}.

The discovery of one or more Higgs boson at present or future colliders will open a new era in the realm of particle
physics. In fact, many efforts have been made to detect such type of particles at colliders.    
In regard to the Higgs
sector of the NMSSM, there has been some work devoted to explore the detectability of at least one Higgs boson at
the LHC and the Tevatron. In particular, some efforts have been made to extend 
the `No-lose theorem' of the MSSM (recall that this
states that at least one Higgs boson of the MSSM will be found at the LHC via the usual SM-like production
and decay channels throughout the entire MSSM parameter space \cite{ATLAS,CMS,NoLoseMSSM}) to the case of the NMSSM 
\cite{Almarashi:2010jm,Almarashi:2011bf,NMSSM-Points,NoLoseNMSSM1,Shobig2,Almarashi:2011hj,Almarashi:2011te,Almarashi:2011qq}. By assuming that 
Higgs-to-Higgs decays are not allowed, it was realised that
at least one Higgs boson of the NMSSM will be discovered at the LHC. However, this theorem could be violated
if Higgs-to-SUSY decays are kinematically allowed (e.g., into neutralino or chargino pairs, yielding 
invisible Higgs signals).

Because of the large number of input parameters of the NMSSM, it is practically very difficult to make a continuous scan over all the 
NMSSM parameter space. The alternative way to do the scan is by resorting to benchmark points in parameter space. 
(For example, for benchmark points in the NMSSM parameter space,
see Refs.~\cite{NMSSM-Points} and \cite{NMSSM-Benchmarks}, for the
unconstrained and constrained case, respectively.) 
Either way, one can distinguish between two scenarios 
in which Higgs-to-Higgs decays are either kinematically allowed or not. 

So far, there is no conclusive evidence
that the `No-lose theorem' can be confirmed in the context of the NMSSM. In order to establish the theorem for the NMSSM, 
Higgs-to-Higgs decays should be taken into account, in particular the decay $h_1\to a_1a_1$. Such a decay can in fact be dominant in large 
regions of the NMSSM parameter space, for instance, for small $A_k$ \cite{Almarashi:2010jm}, and may not give Higgs 
signals with sufficient significance at the LHC. However, a very light CP-odd Higgs boson, $a_1$, can be produced in
association with chargino pairs \cite{Arhrib:2006sx} and in neutralino decays \cite{Cheung:2008rh} at the LHC.

The importance of Higgs-to-Higgs decays in the context of the NMSSM has been emphasised over the years in much literature
in all the above respects, see, e.g., Refs.~\cite{Dermisek:2005ar,Gunion:1996fb,Dobrescu:2000jt,Dobrescu:2000yn}.
 Eventually, it was realised that Vector Boson Fusion (VBF)\footnote{Which is dominated by $W^+W^-$-fusion over $ZZ$-one. }
 could be a viable production channel to detect $h_{1,2}\to a_1a_1$ at the LHC,
 in which the Higgs pair decays into $jj\tau^+\tau^-$ \cite{NMSSM-Points,Ellwanger:2003jt}.
 Some scope could also be afforded by a $4\tau$ signature in both VBF and Higgs-strahlung 
(off gauge bosons) \cite{Belyaev:2008gj}. The gluon-fusion channel too could be a means of
 accessing $h_1\to a_1a_1$ decays, so long that the two light CP-odd Higgs states decay 
into four muons \cite{Belyaev:2010ka} or into two muons and two taus
\cite{2mu2tau}. Such results were all supported by simulations based on parton shower Monte Carlo (MC) programs and
some level of detector response. For a recent survey of the `No-lose theorem' in the NMSSM context, see Ref. \cite{Ellwanger:2011sk}.

Besides, there have also been some attempts to distinguish the NMSSM Higgs sector from the MSSM one, by affirming a 
`More-to-gain theorem' 
\cite{Almarashi:2010jm,Almarashi:2011bf,Almarashi:2011hj,Almarashi:2011te,Almarashi:2011qq,Shobig1,Erice} (that is, to recall, to assess whether 
there exist some areas of the NMSSM parameter space where more and/or different Higgs bosons can be discovered at 
the LHC compared with what is expected from the MSSM). Some comparisons between NMSSM and
MSSM phenomenology, specifically in the Higgs sectors of the two
SUSY realisations,
can be found in \cite{Mahmoudi:2010xp}.

In this paper, we review the LHC discovery potential for the NMSSM Higgs states assuming as production mechanism of these states associated production with bottom-antibottom quark pairs. Generally, the heaviest
CP-even Higgs, $h_3$, the heaviest CP-odd Higgs, $a_2$, and the $h^\pm$ states have very large masses, above the TeV scale, 
in particular at large values
of tan$\beta$, making their discovery at the LHC very difficult. So, we will focus on the LHC discovery potential through this production mode of the lightest CP-odd Higgs boson, $a_1$, and of the lightest two CP-even Higgs states, $h_1$ and $h_2$.


\section{Parameter space scan}
\label{sect:scan1}

As intimated already, due to the large number of parameters in the NMSSM, it is practically not feasible to do a comprehensive scan 
over all of them. These parameters can however be reduced significantly by assuming certain conditions of unification.
Here, since the mechanism of SUSY breaking is still unknown,
to explore the NMSSM Higgs sector, we have performed a general scan in parameter space by fixing the soft 
SUSY breaking terms at high scale to reduce
their contributions to the outputs of the parameter scans. Consequently, we are left with six independent inputs. 
Our parameter space is in particular defined through the Yukawa couplings $\lambda$ and
$\kappa$, the soft trilinear terms $A_\lambda$ and $A_\kappa$ plus 
tan$\beta$ (the ratio of the VEVs of the two Higgs doublets) and $\mu_{\rm eff} = \lambda\langle S\rangle$
(where $\langle S\rangle$, recall, is the VEV of the Higgs singlet). In our numerical analyses we have taken 
$m_b(m_b)=4.214$ GeV, $m_\tau^{\rm pole}=1.777$ GeV,  $m_\mu^{\rm pole}=0.1057$ GeV and $m^{\rm pole}_t=171.4$ GeV 
respectively  
for the running bottom-quark mass and the (pole) tau-lepton, 
muon-lepton and top-quark masses, respectively.

We have used here the {\tt fortran} package NMSSMTools, developed in Refs.~\cite{NMHDECAY,NMSSMTools}\footnote
{We have used NMSSMTools$\_$2.3.1.}. This code
computes the masses, couplings and decay widths of all the Higgs
bosons of the NMSSM, including radiative corrections, in terms of its parameters at the EW
scale. NMSSMTools also takes into account theoretical as well as
experimental constraints from negative Higgs searches at LEP \cite{LEP} and the Tevatron,
including the unconventional channels relevant for the NMSSM.
Notice that the NMSSMTOOLS version used, version 2.3.1,
does not include the latest LHC constraints [29]. However,
as we shall see below, since we keep the SUSY mass
scales very high and, over the phenomenologically interesting
region to this analysis, our $h_1$ state is not very SM-like, the parameter points tested here are safely beyond current LHC limits.

 We have used the  code to scan over the six tree level parameters of the NMSSM Higgs sector
in the following intervals:
\begin{center}
$\lambda$ : 0.0001 -- 0.7,\phantom{aa} $\kappa$ : 0 --
0.65,\phantom{aa} tan$\beta$ : 1.6 -- 54,\\ $\mu_{\rm eff}$ : 100 -- 1000 GeV,\phantom{aa} 
$A_{\lambda}$ : $-$1000 -- +1000 GeV,\phantom{aa} $A_{\kappa}$ :$-$10 -- 0 GeV.\\
\end{center}
\noindent
(Notice that our aim is exploring the parameter space which has very low $m_{a_1}$ and one way to do that is by
choosing $A_\kappa$ small, in which case its negative values are preferred \cite{MNZ}. Also, notice that
small $A_\kappa$ is preferred to have small fine-tuning \cite{excess}.)

Remaining soft terms, contributing at higher order level, which are fixed in the scan include:\\
$\bullet\phantom{a}$ $m_{\tilde{Q}}$ = $m_{\tilde{t}_R}$ = $m_{\tilde{b}_R}$ = $m_{\tilde{L}}$ = $m_{\tilde{\tau}_R}$ = 1 TeV, \\
$\bullet\phantom{a}$ $A_t$ = $A_b$ = $A_{\tau}$ = 1.2 TeV,\\
$\bullet\phantom{a}$ $m_{\tilde{q}}$ = $m_{\tilde{u}_R}$ = $m_{\tilde{d}_R}$ = $m_{\tilde{l}}$ = $m_{\tilde{e}_R}$ = 1 TeV,\\
$\bullet\phantom{a}$ $M_1 = M_2 = M_3 = 1.5$ TeV.\\
As intimated, we have fixed soft term parameters at the TeV scale to minimise their contributions to parameter space
outputs but changing values of some of those parameters such as $A_{t}$ could decrease or increase the number of successful
points emerging from the NMSSMTools scans but without a significant impact on the $m_{a_1}$ distribution.
Also, notice that the sfermion mass parameters and the $SU(2)$ gaugino mass parameter, $M_2$, play 
crucial roles in constraining tan$\beta$. Decreasing values of those parameters allow smaller values of tan$\beta$ to 
pass experimental and theoretical constraints, however, this is a less interesting region of the NMSSM parameter
space for our analysis, as our Higgs production mode is only relevant at large values of tan$\beta$. The effect
of heavy gaugino mass parameters on the outputs, in particular $m_{a_1}$, would be small except
for $M_2$ through its effect on tan$\beta$. In fact, when tan$\beta$ is large, the sfermion masses should be large
to avoid the constraints coming from the muon anomalous magnetic moment \cite{Domingo:2008bb}. The dominant Supersymmetric contribution
at large tan$\beta$ is due to a chargino-sneutrino loop diagram \cite{Czarnecki:2001pv}. Also, notice that the chargino masses
depend strongly on $M_2$.

Guided by the assumptions made in the reference \cite{NMSSM-Points}, the possible decay 
channels for neutral NMSSM CP-even Higgs boson $h$, where $h=h_{1, 2, 3}$, and neutral CP-odd Higgs boson $a$,
where $a=a_{1, 2}$, are:
\begin{eqnarray*}
h,a\rightarrow gg,\phantom{aaa} h,a\rightarrow \mu^+\mu^-,
&&h,a\rightarrow\tau^+\tau^-,\phantom{aaa}h,a\rightarrow
b\bar b,\phantom{aaa}h,a\rightarrow t\bar t, \\ h,a\rightarrow
s\bar s,\phantom{aaa}h,a\rightarrow
c\bar c,&&h\rightarrow W^+W^-,\phantom{aaa}h\rightarrow ZZ, \\
h,a\rightarrow\gamma\gamma,\phantom{aaa}h,a\rightarrow
Z\gamma,&&h,a\rightarrow {\rm Higgses},\phantom{aaa}h,a\rightarrow
{\rm sparticles}.
\end{eqnarray*}
(Notice that the CP-odd Higgses are not allowed to decay into vector
boson pairs due to CP-conservation.) Also, notice that here
`${\rm Higgses}$' denotes any possible final state
involving two neutral or two charged Higgs bosons or one Higgs boson and one gauge boson.

We have performed a random scan over millions of points in the specified parameter space. The output of the scan, as
mentioned above, contains masses, Br's and couplings of
the NMSSM Higgses for all the successful points which have passed 
the various experimental and theoretical constraints.

\section{Inclusive event rates}
\label{sect:rates1}

For the successful data points, we used CalcHEP \cite{CalcHEP} to calculate the
cross sections for NMSSM Higgs production\footnote{We adopt herein
CTEQ6L \cite{cteq} as parton distribution functions, with scale $Q=\sqrt{\hat{s}}$, the centre-of-mass energy
at parton level, for all processes computed.}. Some new modules have been implemented for this purpose. 

We focus here on the process 
\begin{equation}
gg\to b\bar b~{a_1} 
\label{eq:proc}
\end{equation}
i.e., Higgs production in association with a $b$-quark pair. (The production mode $q\bar q\to b\bar b~{a_1}$ is negligible at the LHC
with $\sqrt{s}$ = 14 TeV.)
We chose the production mode $gg\to b\bar ba_1$ because it is the dominant one at large tan$\beta$. The gluon
fusion channel is instead burdened by huge SM backgrounds and $a_1$ does not couple to gauge bosons
in Higgs-strahlung and Vector Boson Fusion (VBF) processes due to CP-conservation, see \cite{Almarashi-talk}. In addition, Higgs
production in other modes has been studied before, see for example \cite{Shobig2}. 
In fact, Higgs production in association with a $b\bar b$ pair has an extra advantage, whereby the associated $b\bar b$ pair can
be tagged, allowing a useful handle for background rejection. 

\begin{figure}[h]
  \centering\begin{tabular}{cc}
 \includegraphics[scale=1]{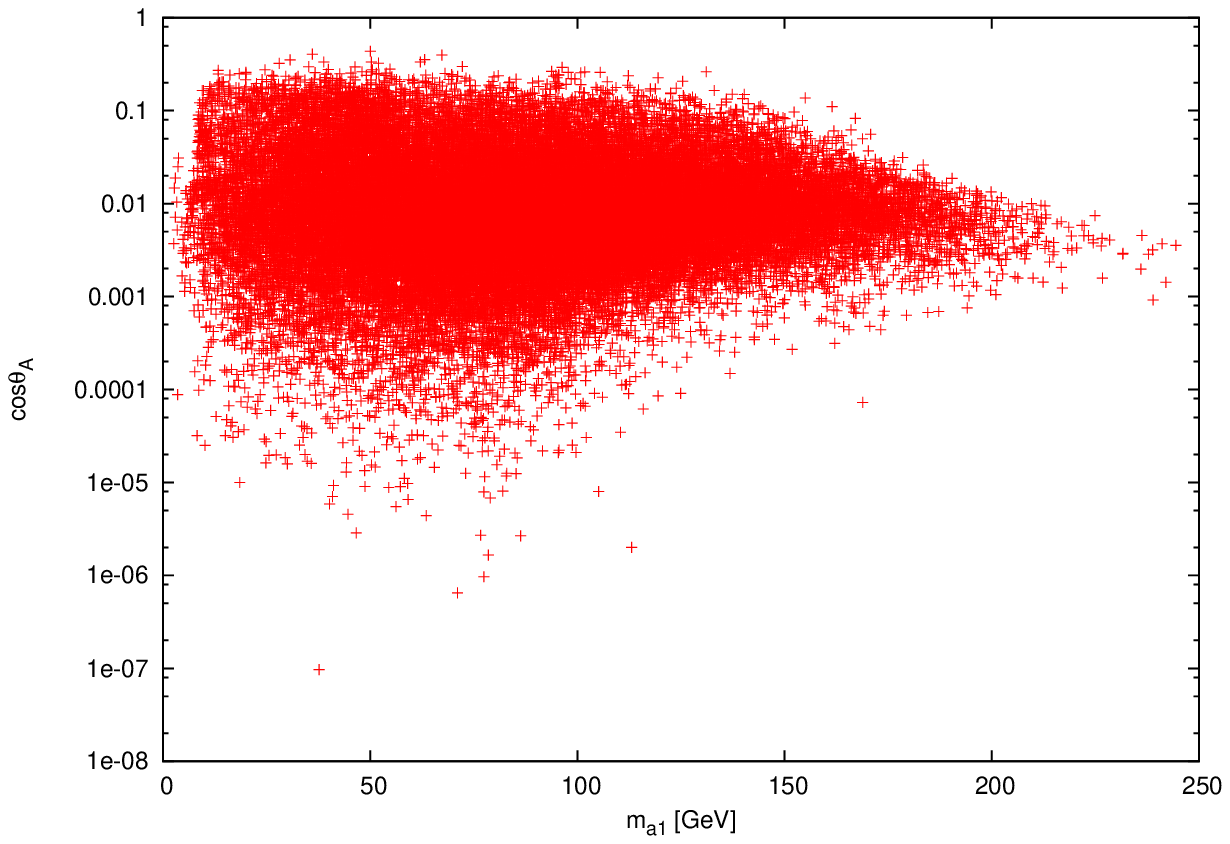}\\
\includegraphics[scale=1]{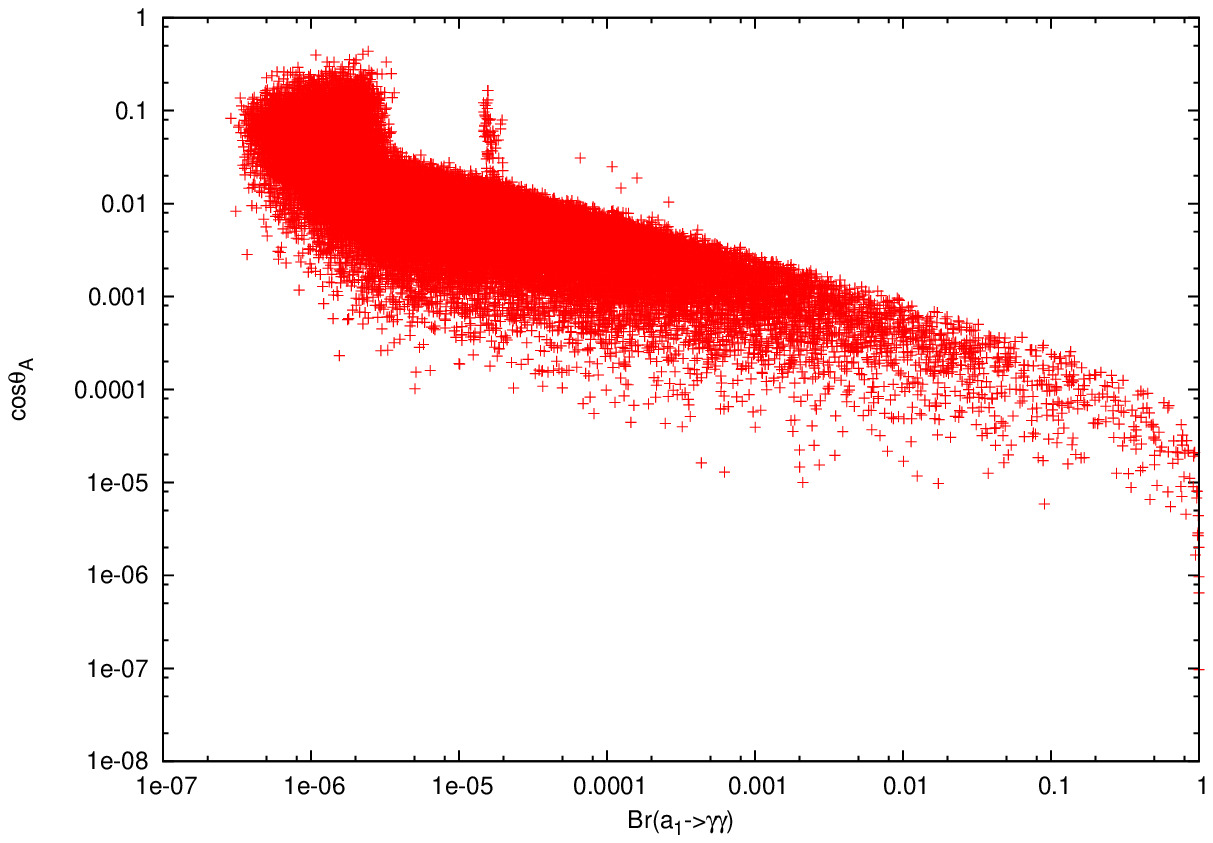}
 \end{tabular}
\caption{The lightest CP-odd Higgs mass $m_{a_1}$ and the Br$(a_1\to \gamma\gamma)$ plotted against the mixing angle
in the CP-odd Higgs sector $\cos\theta_A$. }
\label{fig:costheta}
\end{figure}

In the NMSSM, the $a_1$ state is a composition of the usual doublet component of the CP-odd MSSM Higgs boson, $a_{\rm MSSM}$, and
the new singlet component, $a_{\rm S}$, coming from the singlet Superfield of the NMSSM. This can be written as \cite{excess}:
\begin{equation}
 a_1=a_{\rm MSSM}\cos\theta_A+a_{\rm S}\sin\theta_A.
\end{equation}
For very small values of $A_k$, the lightest CP-odd Higgs, $a_1$, is mostly singlet-like
with a tiny doublet component, i.e., the mixing angle $\cos\theta_A$ is small, see
the top-pane of figure~\ref{fig:costheta} which shows the relation between $m_{a_1}$ and $\cos\theta_A$. 
The bottom-pane of the figure shows that the Br$(a_1\to \gamma\gamma)$ can be dominant in some regions of the NMSSM parameter space with
the possibility of reaching unity when $\cos\theta_A\sim 0$.

To a good approximation, $m_{a_1}$ can be written in the NMSSM as \cite{excess}:
\begin{equation} 
 m^2_{a_1}=-3\frac{\kappa A_\kappa \mu_{\rm eff}}{\lambda}\sin^2\theta_A+\frac{9A_\lambda \mu_{\rm eff}}{2\sin2\beta}\cos^2\theta_A.
\end{equation}
The first term of this expression is dominant at large tan$\beta$. Furthermore, it is clear that
a combination of all the tree level Higgs sector parameters affects $m_{a_1}$ in general.

\section{\bf Photon and tauon signals of very light CP-odd Higgs states of the NMSSM at the LHC}

In our attempt to test the two aforementioned theorems, we consider in this section the case of the $\gamma\gamma$ and $\tau^+\tau^-$ decay channels
of a very light CP-odd Higgs boson. The first mode is
the most important one to detect a CP-even Higgs boson below 130 GeV in the SM and MSSM despite the smallness of its 
branching ratio, of $\cal{O}$$(0.001)$. In addition, this decay mode gives a clean signature and can be resolved 
efficiently at the LHC. The second one is used in the MSSM as a search channel of rather heavy 
CP-even and CP-odd states, in particular at large tan$\beta$, and its exploitation has not been proved at very low masses,
say, below $M_Z$.

In the NMSSM, because of the introduction of a complex singlet Superfield, the lightest CP-odd Higgs boson, $a_1$, can be
a singlet-like state with a tiny doublet component in large regions of parameter space. In this section
(and also in the next one) we are looking for direct production of the
$a_1$ rather than looking for its traditional production through $h_{1, 2}$ decay. We examine
the discovery potential of the $a_1$ produced in association with a bottom-antibottom pair at the LHC through the
$\gamma\gamma$ and $\tau^+\tau^-$ decay modes. 

We will show that in the NMSSM there exist regions of its parameter
space where one can potentially have a dominant di-photon branching ratio of $\cal{O}$$(1)$ 
for the lightest CP-odd Higgs boson with small mass. This possibility emerges in
the NMSSM because of the fact that such a CP-odd Higgs state has a predominant 
singlet component and a very weak doublet one. As a
consequence, all partial decay widths are heavily suppressed as they
employ only the doublet component, except one: the $\gamma\gamma$ partial decay
width. This comes from the fact that the $a_1\tilde\chi^+\tilde\chi^-$
coupling is not suppressed, as it is generated through the $\lambda H_1 H_2 S$ Lagrangian
term and therefore implies no small mixing. Although the direct decay 
$a_1$ $\to$ $\tilde\chi^+\tilde\chi^-$ is kinematically not allowed, the aforementioned
coupling participates in the $a_1\gamma\gamma$ effective coupling \cite{communication}. 

Furthermore, we will show that the $\tau$-pair decay can be a promising decay mode for detecting the $a_1$ state of the NMSSM
with very low mass. The detection of such a very low mass Higgs state would then unmistakably
signal the existence of a non-minimal SUSY Higgs sector.

\begin{figure} 
 \centering\begin{tabular}{cc}
\includegraphics[scale=0.6]{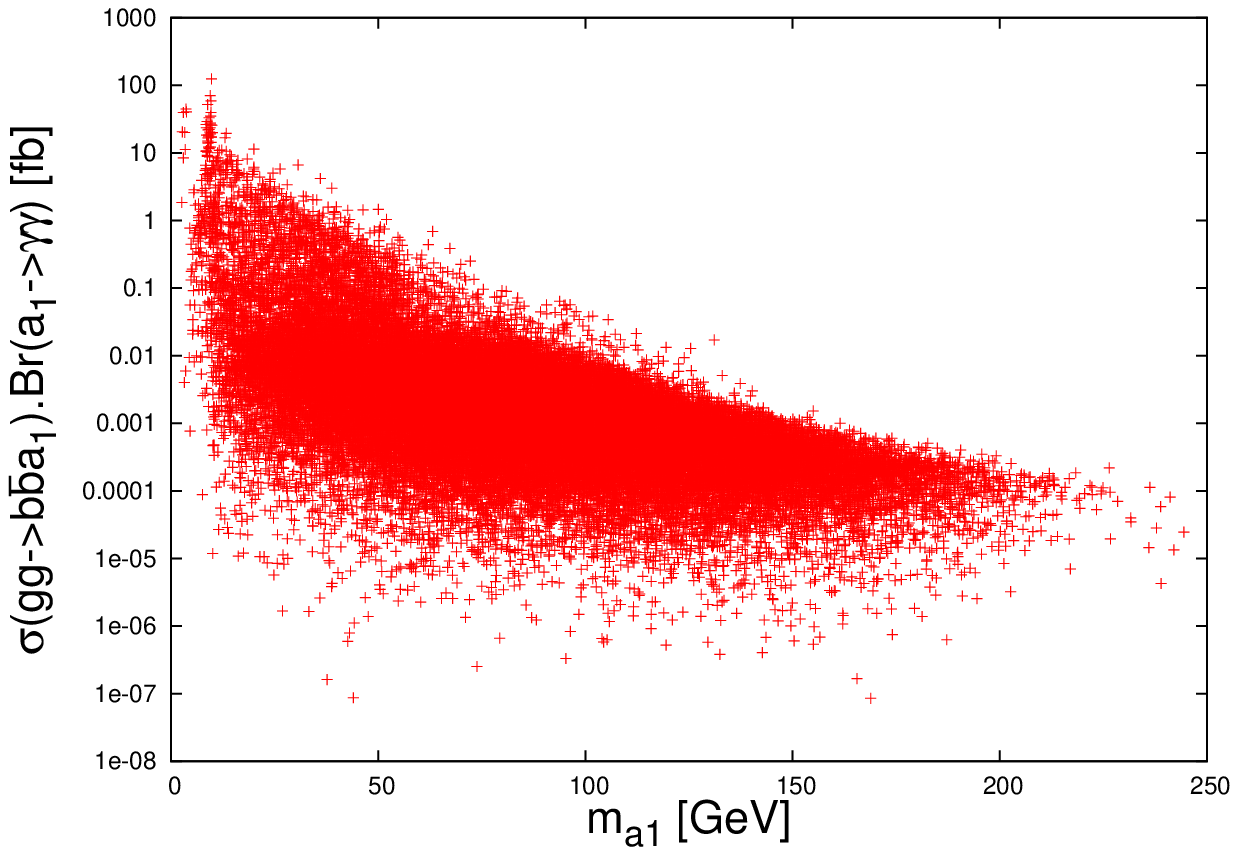}
&\includegraphics[scale=0.6]{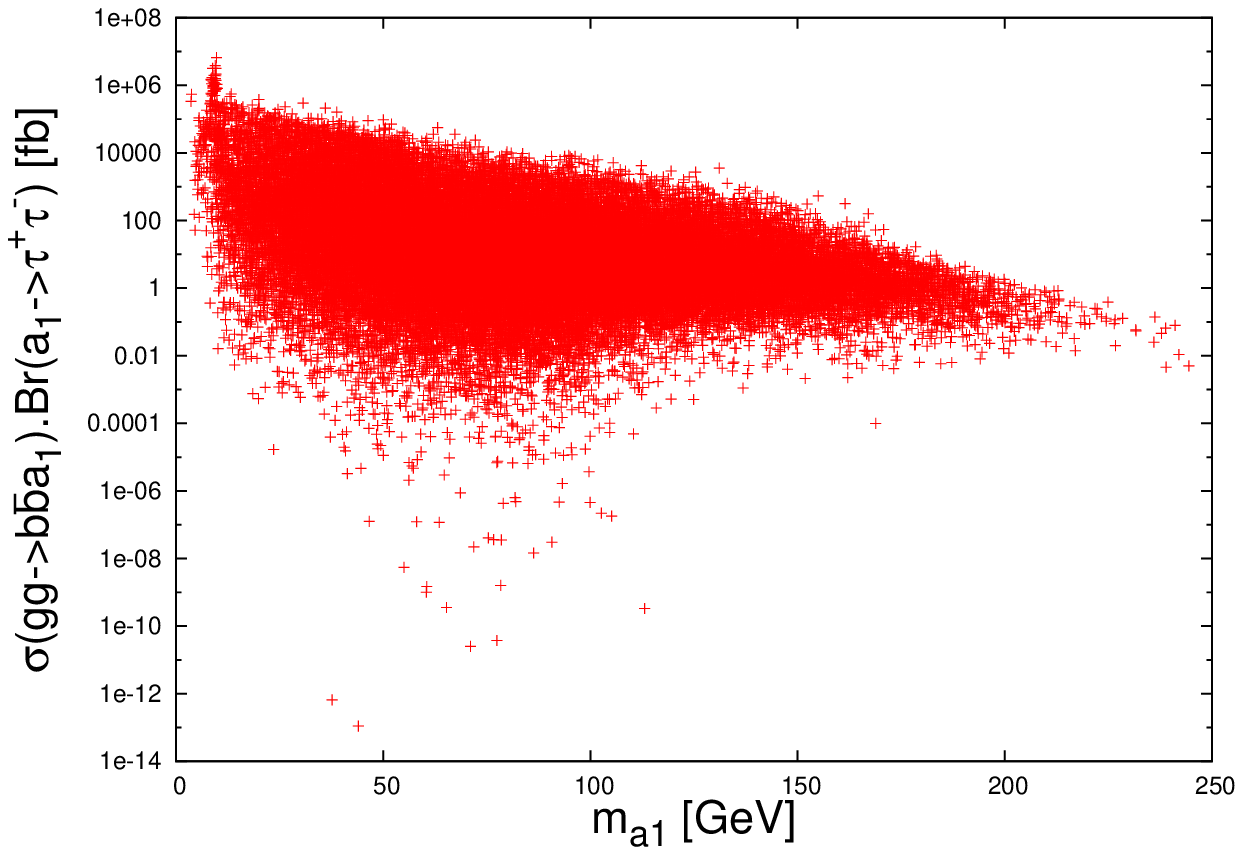}\\
\includegraphics[scale=0.6]{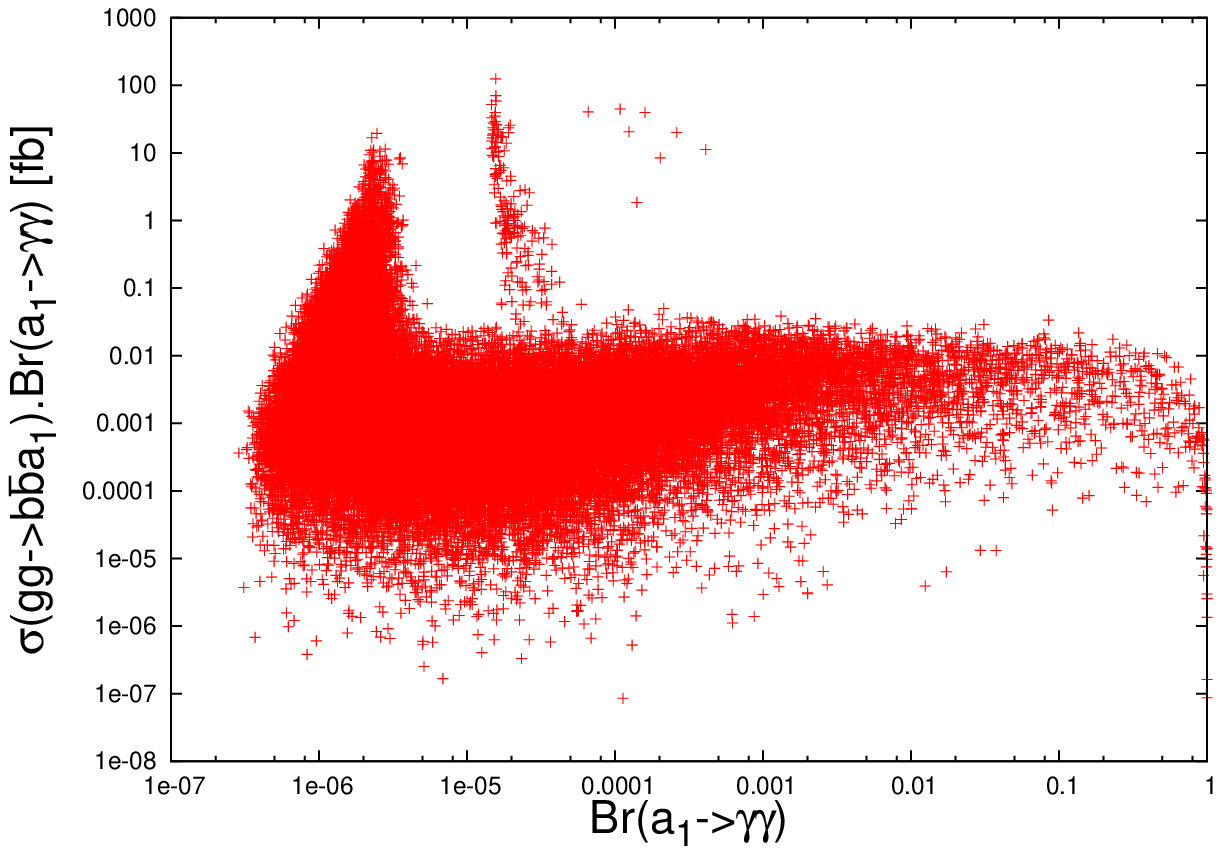}
&\includegraphics[scale=0.6]{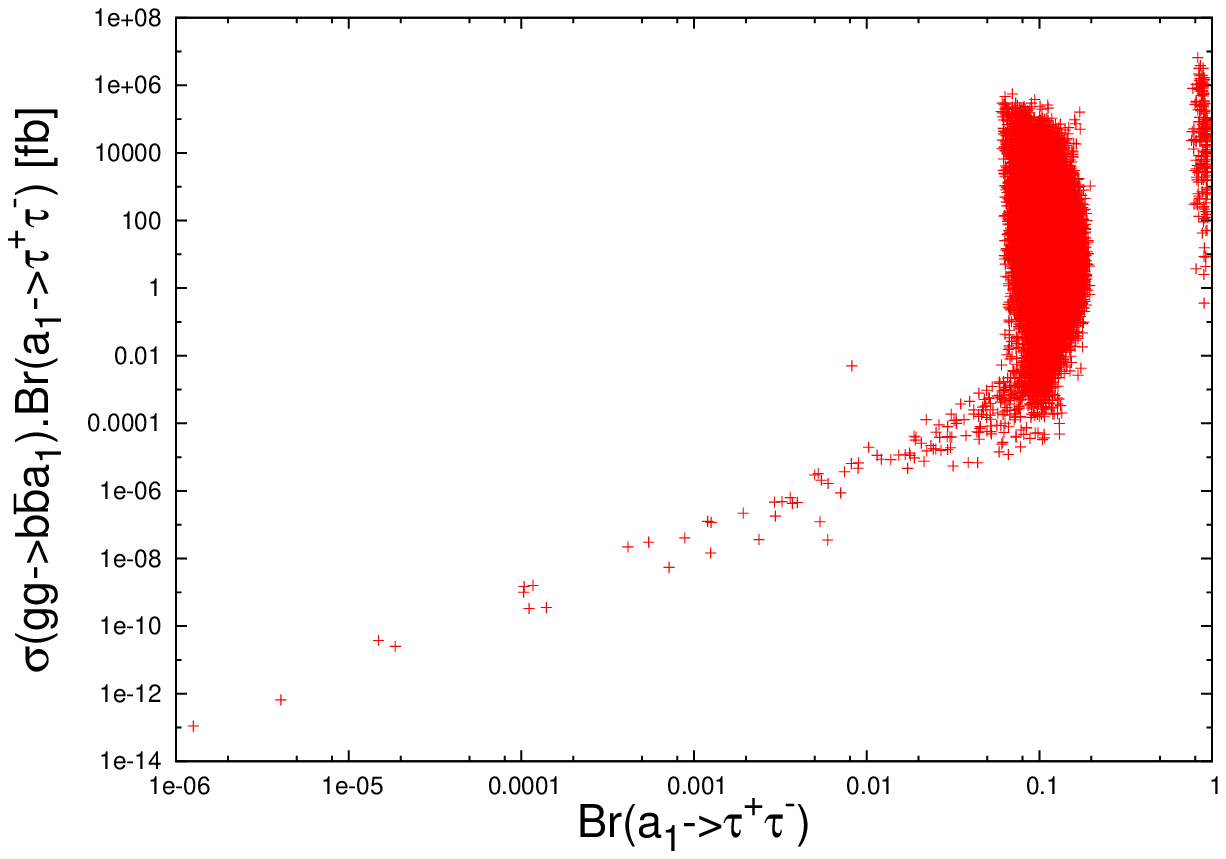}\\

 \end{tabular}
\caption{The rates for $\sigma(gg\to b\bar b {a_1})~{\rm Br}(a_1\to \gamma\gamma)$ (left) and
for $\sigma(gg\to b\bar b {a_1})~{\rm Br}(a_1\to \tau^+\tau^-)$ (right) as functions of $m_{a_1}$ and of the Br of the corresponding channel. }
\label{fig:sigma-scanAA5-3}
\end{figure}

Figure~\ref{fig:sigma-scanAA5-3} shows the distribution of the event rates $\sigma(gg\to b\bar b {a_1})~{\rm Br}(a_1\to \gamma\gamma)$ and
$\sigma(gg\to b\bar b {a_1})~{\rm Br}(a_1\to \tau^+\tau^-)$ as functions of $m_{a_1}$ and of Br's of the corresponding channel. 
As expected, the inclusive cross section decreases with increasing $m_{a_1}$, 
see the top panes of the figure. It is worth mentioning that the Br$(a_1\to \gamma\gamma)$ can be dominant over a sizable expanse of
the NMSSM parameter space, which originates from tiny widths into all other channels due to the dominant
singlet nature of $a_1$ as mentioned in Sec. 5.1. However,
the dominance of Br$(a_1\to \gamma\gamma)$  
does not correspond to the region that maximises the yield of $\sigma(gg\to b\bar b {a_1})~{\rm Br}(a_1\to \gamma\gamma)$, 
as the maximum
of the latter occurs for Br's in the region of some $10^{-5}$ to $10^{-4}$, see the bottom-left pane of the figure. 
Therefore, one can not 
take full advantage of the phenomenon described in the introduction of this section with respect to the singlet nature of the
$a_1$ state, at the LHC, which couples to $\gamma\gamma$ through charginos. Thus, if $a_1$ were highly singlet, it would be
difficult for the LHC to discover this particle as the doublet component (necessary to enable a 
large $a_1b\bar b$ coupling at production level) would be suppressed.
The tension between the two components 
 is such that the cross section times Br rates are less than 100 $fb$.

The outlook for the $\tau^+\tau^-$ decay mode 
is much brighter where the corresponding signal rates are at $nb$ level for 
Br$(a_1\to \tau^+\tau^-)\approx 0.1$ or even 10 $nb$ for Br$(a_1\to \tau^+\tau^-)\approx 1$, see the bottom-right
pane of figure~\ref{fig:sigma-scanAA5-3}. Also, notice that such large rates naturally hold for different values of $m_{a_1}$, 
in the allowed interval, but they decrease with increasing $m_{a_1}$ (see the top-right pane of this figure).

\begin{figure} 
 \centering\begin{tabular}{cc}
\includegraphics[scale=1]{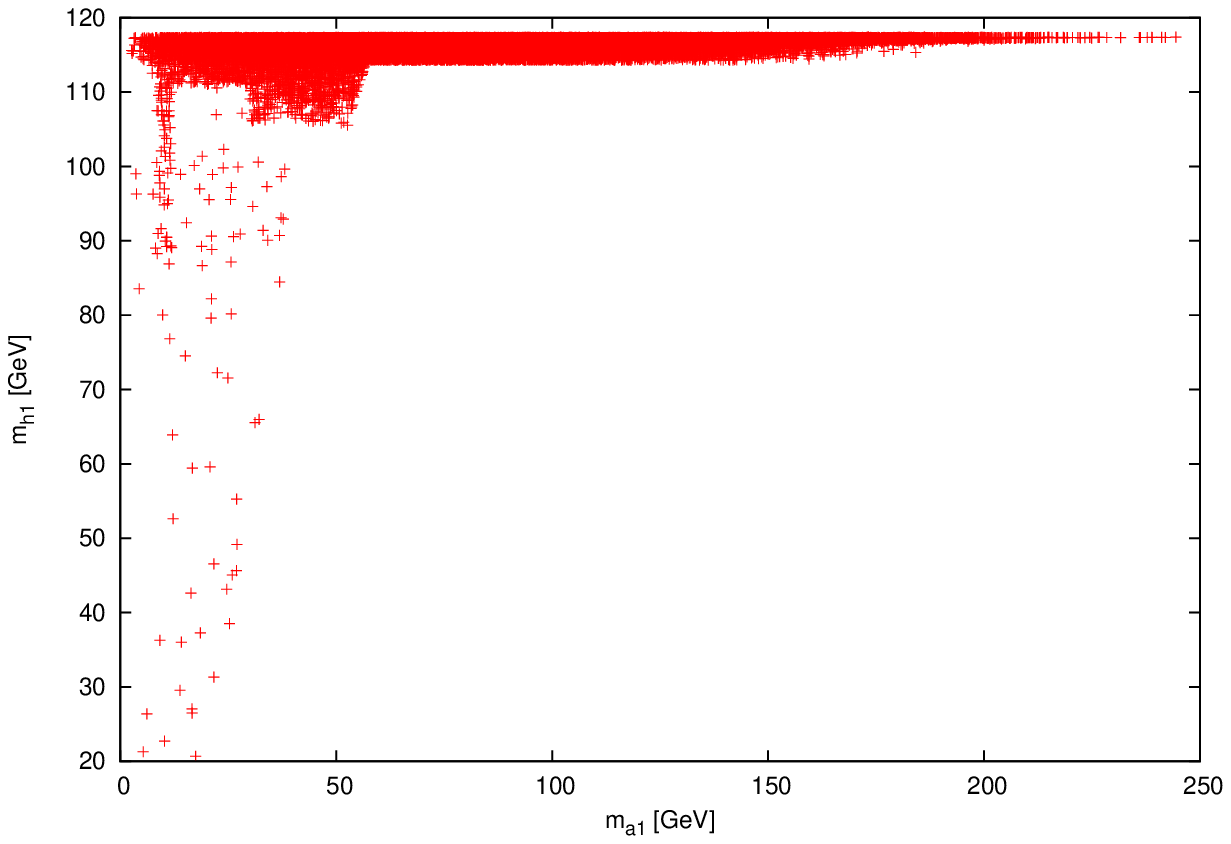}\\
\includegraphics[scale=1]{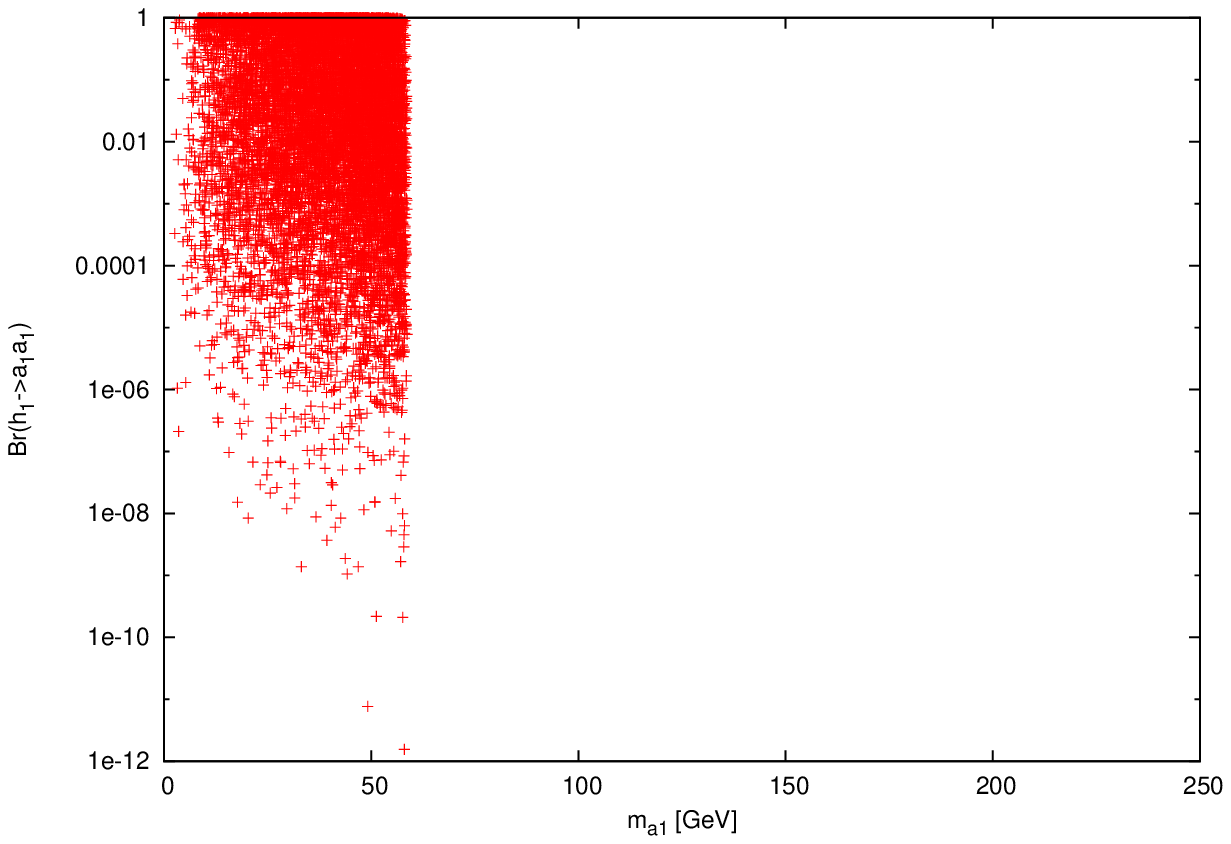}
 \end{tabular}
\caption{The lightest CP-odd Higgs mass $m_{a_1}$ plotted against the lightest CP-even Higgs mass $m_{h_1}$
 and against Br$(h_1\to a_1a_1)$. }
\label{fig:mh1-ma15-4}
\end{figure}

In the NMSSM, there is a large area of parameter space where one Higgs state can decay into two, e.g., $h_1\to a_1a_1$: see
figure~\ref{fig:mh1-ma15-4}.  As it is clear from the top-pane of this figure, the majority of points generated here have $m_{h_1}>110$ GeV and 
$m_{a_1}<55$ GeV, thereby allowing the possibility of $h_1\to a_1a_1$ decays. Moreover, this decay can be dominant and can reach unity as 
shown in the bottom-pane of the figure. Despite this, such a decay may not give Higgs signals with sufficient statistical significance 
at the LHC (as discussed in previous literature). Therefore, we are well motivated to study, in the fortchomin sections, the scope of direct
production of the $a_1$ state in single mode at the LHC, through $gg\to b\bar ba_1$, over overlapping
regions of NMSSM parameter space{\footnote{A partonic signal-to-background
($S/B$) analysis for $\gamma\gamma$ and $\tau^+\tau^-$ final states has been done in \cite{Almarashi:2010jm}, where extraction of the latter signature was proven for several benchmark scenarios.}}.

\section{\bf Muon and $b$-quark signals of very light CP-odd Higgs states of the NMSSM at the LHC}
\label{sect:intro}

The di-muon decay mode has an advantage, that it has a clean signature with excellent mass resolution. However, the $\mu^+\mu^-$ branching
ratio is small in most regions of parameter space but this decay mode is enhanced for large tan$\beta$.

\begin{figure}
 \centering\begin{tabular}{c}
  \includegraphics[scale=1]{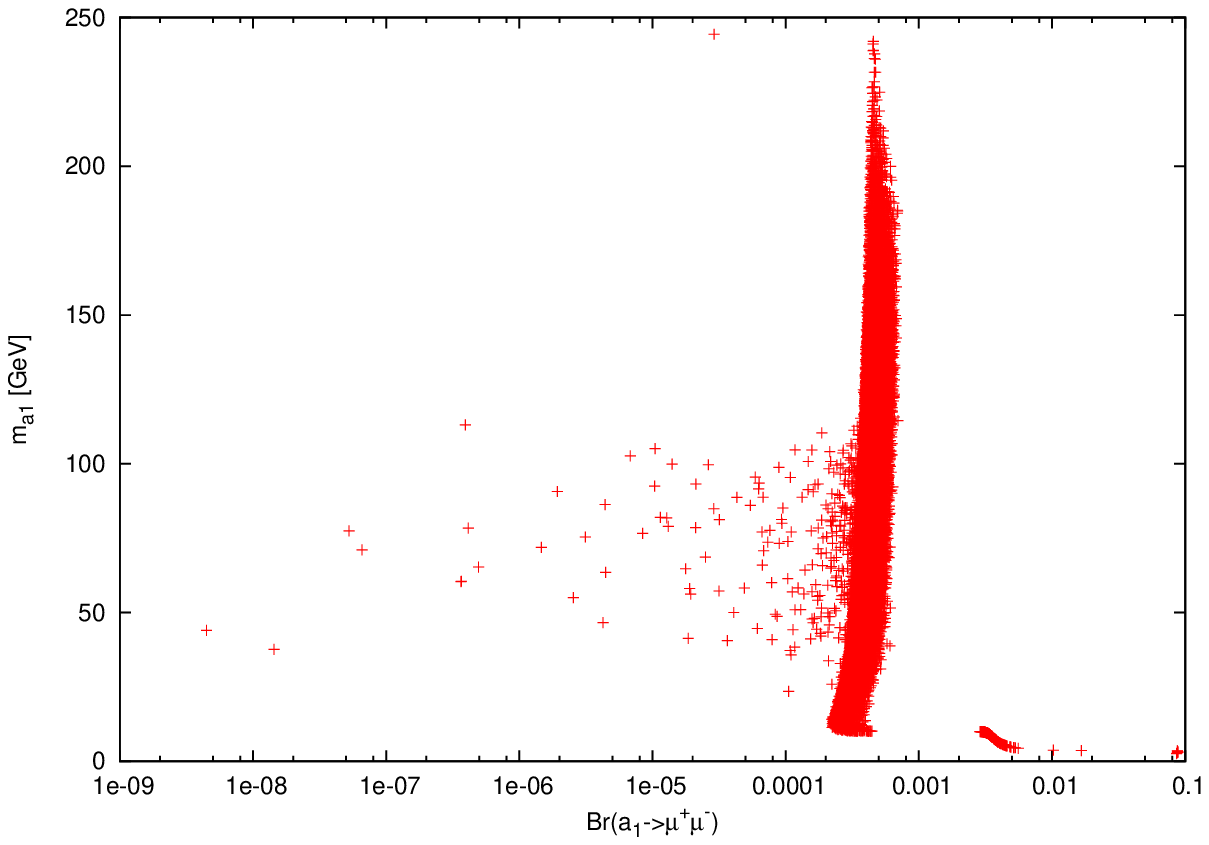}
 \end{tabular}
\caption{The CP-odd Higgs mass $m_{a_1}$ as a function of the Br$(a_1\to \mu^+\mu^-)$.}
\label{fig:mass-scan2mu}
\end{figure}

Figure \ref{fig:mass-scan2mu} shows the correlations between the 
$a_1$ mass and the di-muon decay rate. One can see from this figure
that the Br$(a_1\to\mu^+\mu^-)$ can be of ${\cal O}(10\%)$, ${\cal O}(1\%)$ and ${\cal O}(0.1\%)$ or less for the mass intervals 
$2m_\mu<m_{a_1}<2m_\tau$, $2m_\tau<m_{a_1}<2m_b$ and $2m_b<m_{a_1}$, respectively. The first region of parameter space 
($m_{a_1}<2m_\tau$) is rather small, the second one
($2m_\tau<m_{a_1}<2m_b$) more significant and the third one ($2m_b<m_{a_1}$) is by far the widest one.

\begin{figure}
 \centering\begin{tabular}{cc}
 \includegraphics[scale=0.58]{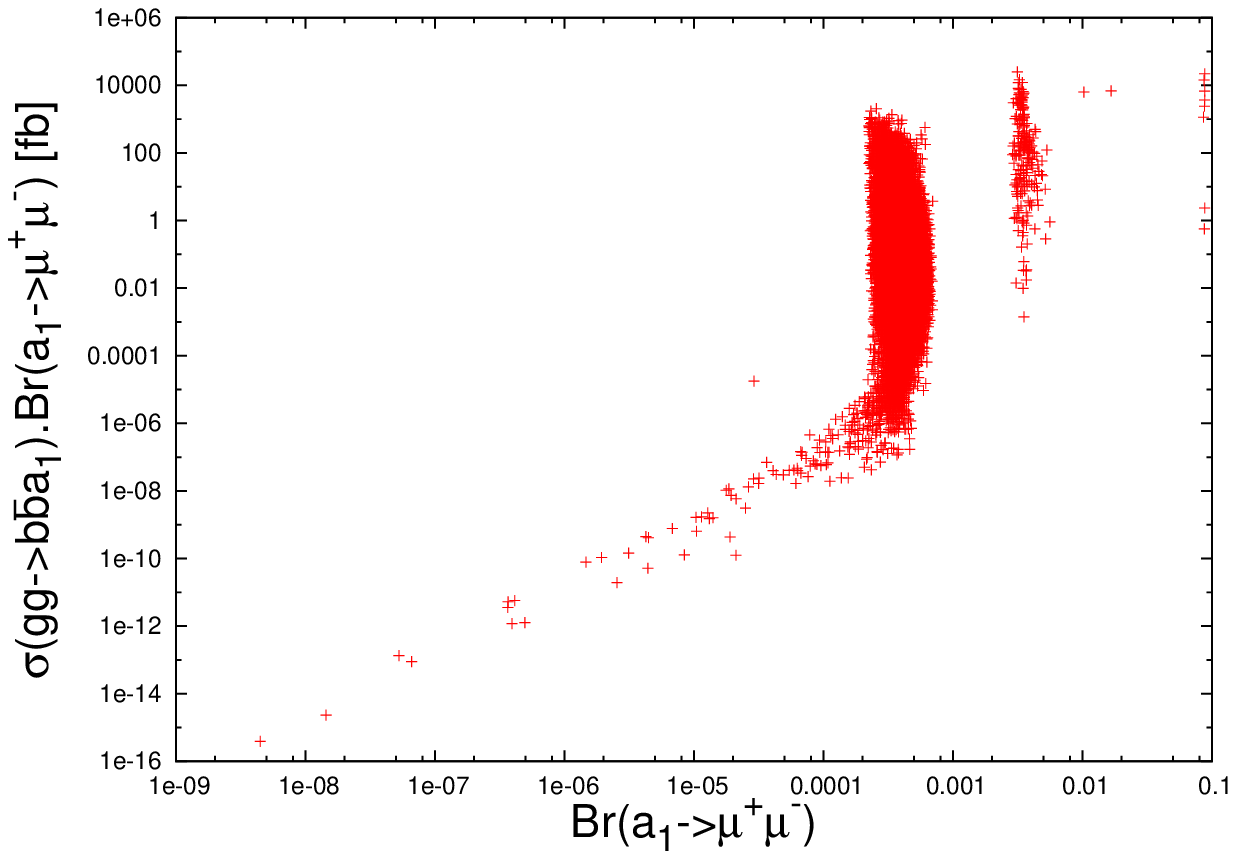}&\includegraphics[scale=0.58]{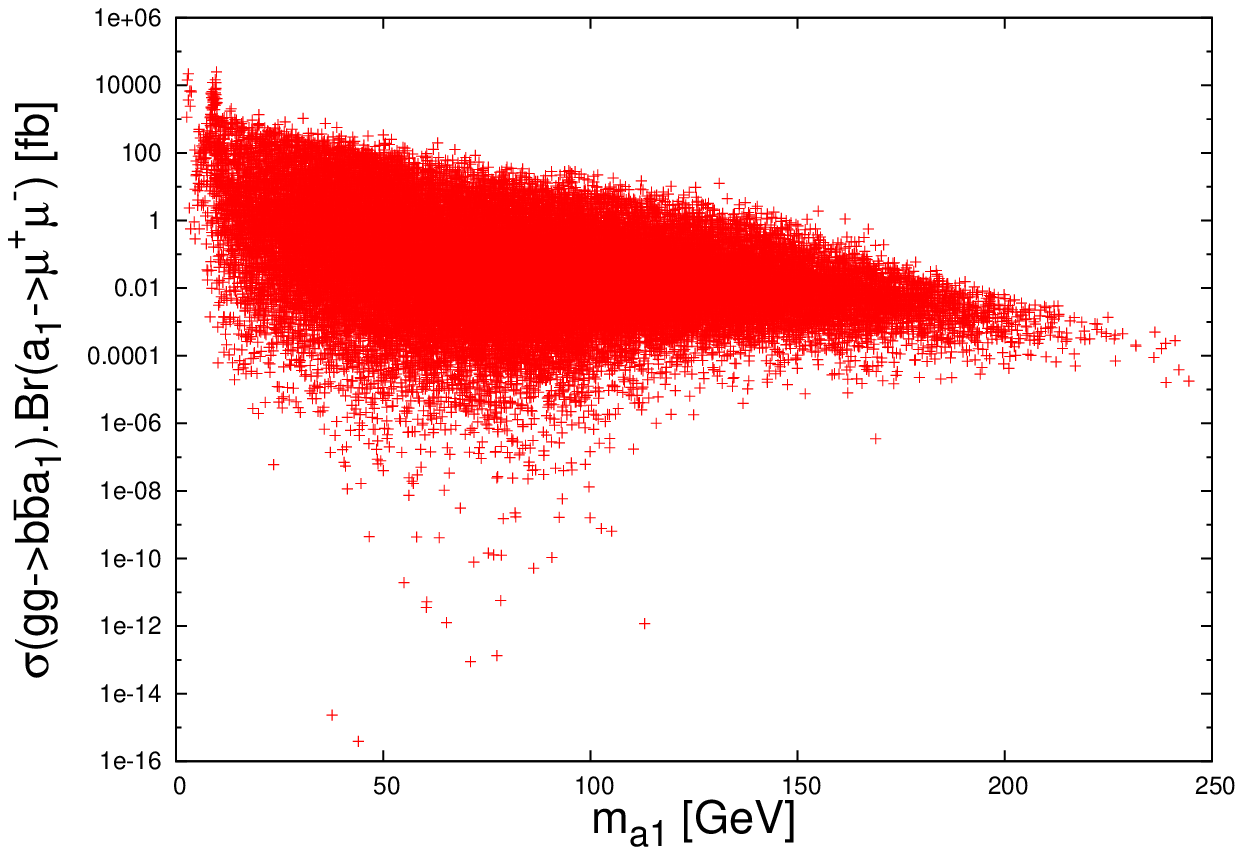}
 \end{tabular}
\caption{The rates for $\sigma(gg\to b\bar b {a_1})~{\rm Br}(a_1\to \mu^+\mu^-)$ as a function of  ${\rm Br}(a_1\to \mu^+\mu^-)$ and of $m_{a_1}$. }
\label{fig:sigma-scan2muon}
\end{figure}

Figure \ref{fig:sigma-scan2muon} illustrates the distribution
of the inclusive event rates as a function of the Br$(a_1\to \mu^+\mu^-)$
and of $m_{a_1}$. It is remarkable to notice that the inclusive event rates are sizable in all
such mass regions. These event rates reach the
$10^4$ $fb$ level in the two lower mass intervals and the $10^3$ $fb$ level in the higher mass range and clearly decrease by
increasing $m_{a_1}$, as expected. Finally, notice that the mass
region below the $\mu^+\mu^-$ threshold is very severely constrained \cite{Lebedev}. 

As for $4b$-quark final states, at large tan$\beta$ values, the cross-section of the $a_1$ produced in association with a bottom-antibottom pair
followed by the decay $a_1\to b\bar b$ is strongly enhanced, in general. 
However, since the channel is a 4-quark final state, it is plagued by very
large irreducible and reducible backgrounds. In this section, we examine whether or not the production mode 
$gg\to b\bar ba_1\to b\bar b b\bar b$ can be exploited to detect the $a_1$ at the LHC.
In fact, the existence of $b$-jets in the final states offers the advantage of $b$-tagging, which can be exploited to trigger 
on the signal and enable us to require up to four displaced vertices in order to reject light jets.
The ensuing $4b$ signature has already been exploited to detect neutral Higgs bosons of the
 MSSM at the LHC and proved useful, provided that tan$\beta$ is large and the collider has good 
efficiency and purity in tagging $b$-quark jets, albeit
for the case of rather heavy Higgs states (with masses beyond $M_Z$, typically)
\cite{Dai:1994vu,Dai:1996rn}. 

\begin{figure}
 \centering\begin{tabular}{cc}
     \includegraphics[scale=0.60]{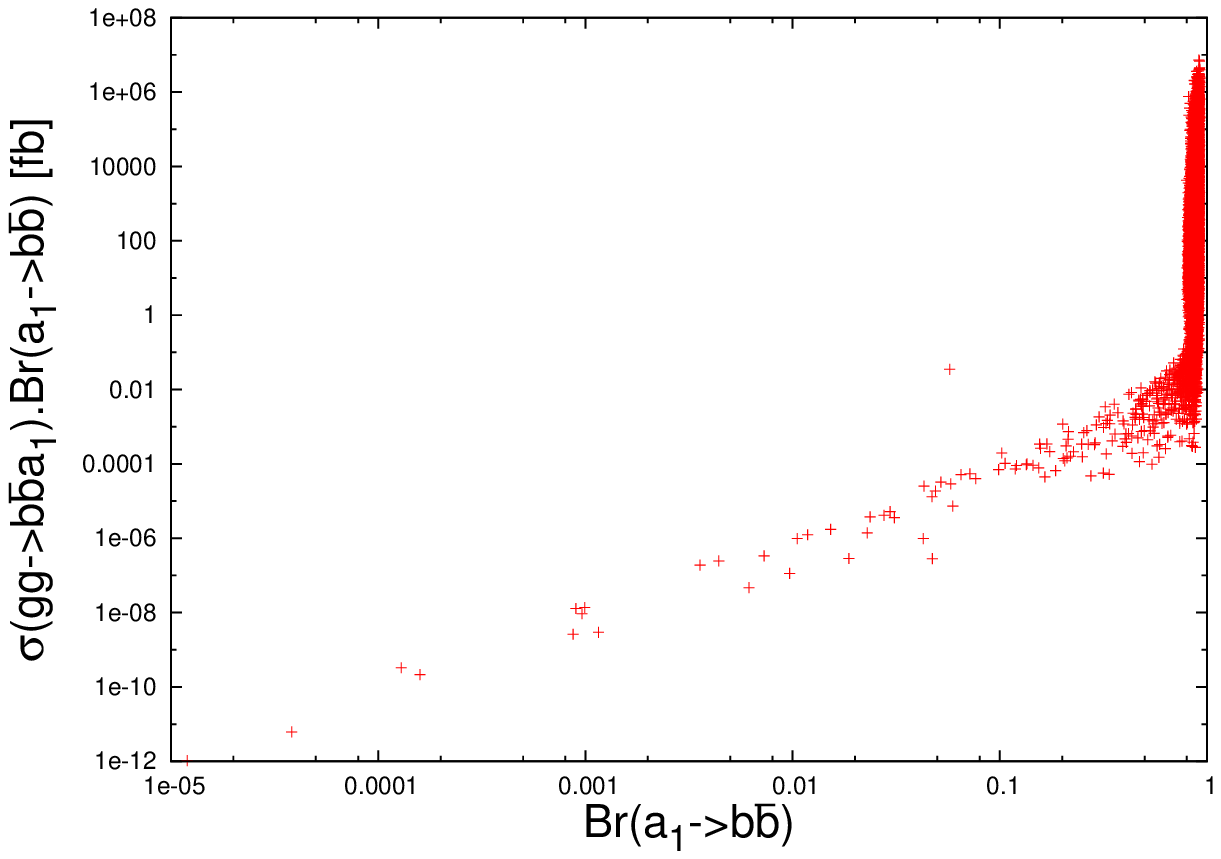}&\includegraphics[scale=0.60]{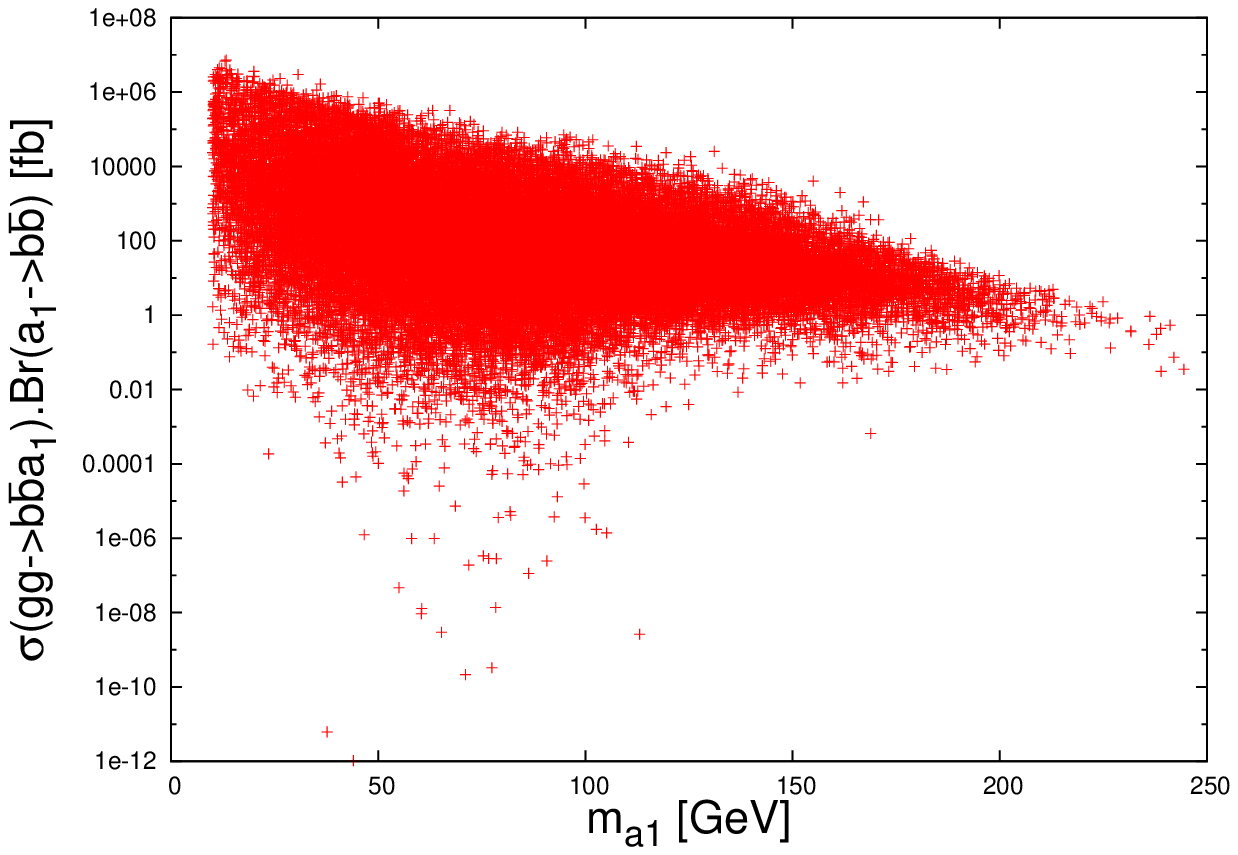}
 \end{tabular}
\caption{The rates for $\sigma(gg\to b\bar b {a_1})~{\rm Br}(a_1\to b\bar b)$ as a function of the Br$(a_1\to b\bar b)$ and of $m_{a_1}$. }
\label{fig:bBscan}
\end{figure}

\begin{figure}
 \centering\begin{tabular}{cc}
 
   \includegraphics[scale=1]{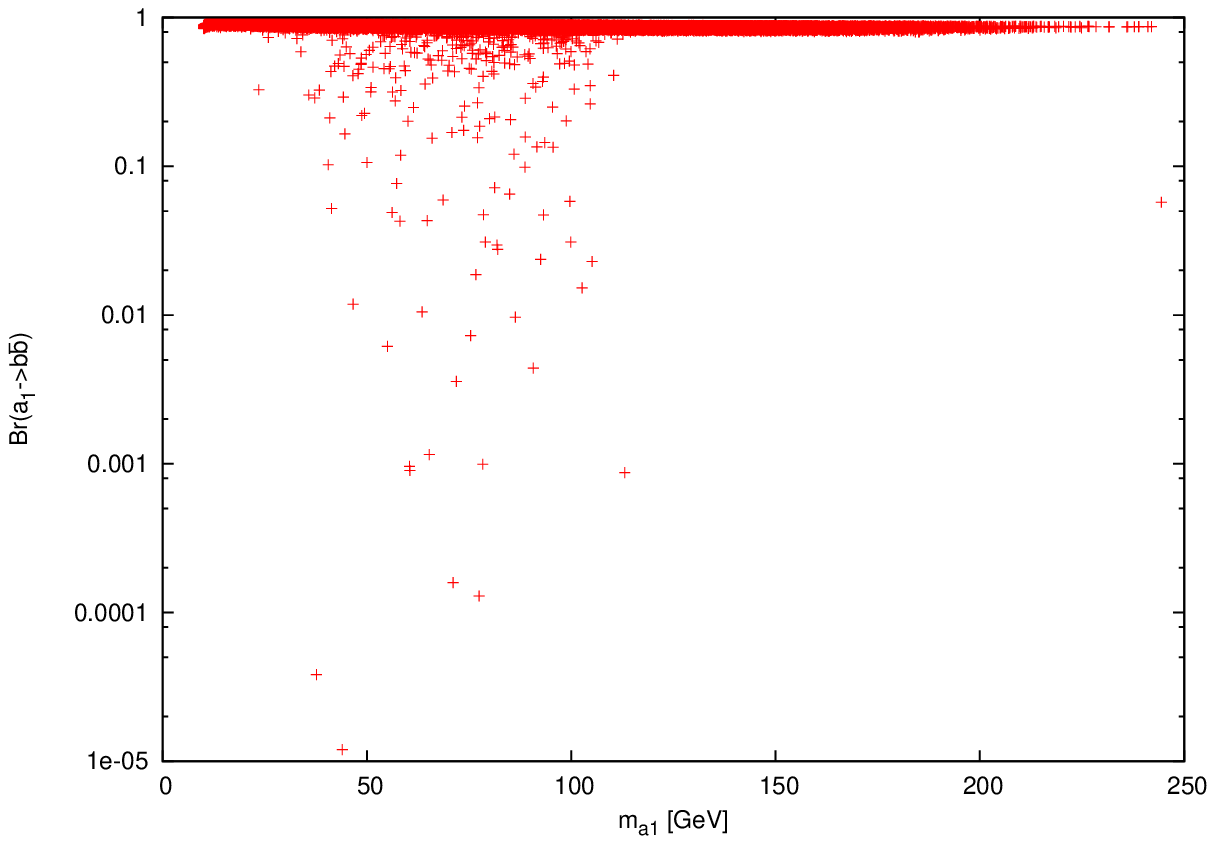}\\
\includegraphics[scale=1]{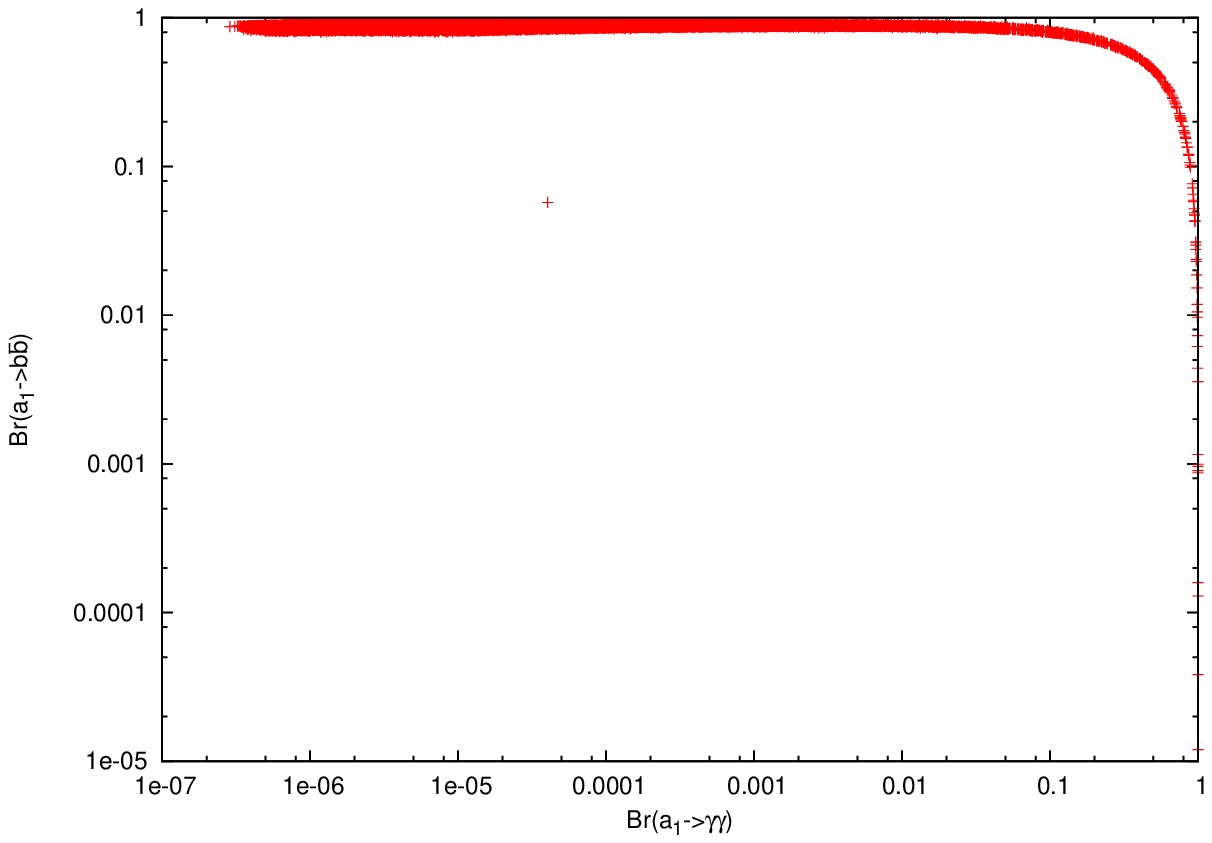}

 \end{tabular}
\caption{The Br$(a_1\to b\bar b)$ as a function of the CP-odd Higgs mass $m_{a_1}$ and of the Br$(a_1\to \gamma\gamma)$ .}
\label{fig:bBmass-branchingratio}
\end{figure}

Figure \ref{fig:bBscan} illustrates the inclusive signal production  
cross section $\sigma(gg\to b\bar ba_1)$ multiplied by the branching fraction Br$(a_1\to b\bar b)$ 
as a function of the Br$(a_1\to b\bar b)$ and of $m_{a_1}$ and 
the plots in figure \ref{fig:bBmass-branchingratio} display instead the correlations between the
$a_1\to b\bar b$ decay rate and the $a_1$ mass (top-pane) and between the $a_1\to b\bar b$ decay rate 
and the $a_1\to\gamma\gamma$ decay rate. From a close look at the bottom-left pane
of figure \ref{fig:bBscan}, it is clear 
that the Br$(a_1\to b\bar b)$ is dominant for most points in the parameter space, about 90\% and above.
In addition, by looking at the the bottom-right pane of the figure, it is remarkable to notice that also these event rates are sizable
in most regions of parameter space, topping the
$10^7$ $fb$ level for small values of $m_{a_1}$ and decreasing rapidly with increasing $m_{a_1}$. One
can also notice that there are
some points in the parameter space with $m_{a_1}$ between 40 and 120 GeV, as shown in the top-pane of figure \ref{fig:bBmass-branchingratio},
in which the Br$(a_1\to b\bar b)$ is suppressed due to the enhancement of the Br$(a_1\to \gamma\gamma)$ 
(see the bottom-pane of the same figure), a phenomenon peculiar to the NMSSM that depends upon the amount
of Higgs singlet-doublet mixing\footnote{Notice that constraints coming from Tevatron \cite{Abazov:2011jh}
do not affect our results
since the singlet field plays a primary role in the NMSSM. But under severe conditions such as $\lambda\to 0$ 
and $\kappa\to 0$,
the NMSSM and MSSM become similar and those constraints may be applied.}, see \cite{Almarashi:2010jm}. (We discussed this in the previous
section{\footnote{A partonic signal-to-background
($S/B$) analysis for $\mu^+\mu^-$ and $b\bar b$ final states has been done in \cite{Almarashi:2011bf,Almarashi:2011hj}, extracting both signals for several benchmark scenarios.}}.) 

\section{\bf 
 The `No-lose theorem' for NMSSM Higgs discovery at the LHC in difficult scenarios}

In this section, we continue to investigate whether or not the `No-lose theorem' of
the NMSSM at the LHC can be proven considering Higgs boson production in association with a $b$-quark pair. We do so based on the interesting results obtained in the previous
two sections. In this section we will, however, no longer consider direct $a_1$ production, i.e., $gg\to b\bar b a_1$. Rather, we will initially produce either a $h_1$ or $h_2$, eventually decaying to one or more $a_1$'s. In this case one may wonder though whether also Higgs boson production in association with a $t$-quark pair could play a role, owning to different couplings of the $h_{1,2}$ state 
to fermions, with respect to the $a_1$ field. Production rates for
$gg\to t\bar t h_{1,2}$ were studied in \cite{Shobig2},
where they were found to be very subleading over the entire NMSSM parameter space. 

We will be looking at inclusive event rates in presence of  
various Higgs-to-Higgs decays, namely, $h_{1, 2}\to a_1a_1$, $h_2\to h_1h_1$ and $h_{1, 2}\to Za_1$. We will also be studying the decay patterns of the lightest Higgs boson pairs, 
$a_1a_1$ or $h_1h_1$, and of the gauge boson and a light CP-odd Higgs boson, $Za_1$, into different final states. Further details on material contained in this section can be found in \cite{Almarashi:2011te} and 
\cite{Almarashi:2011qq}.

We have again used NMSSMTools to perform a random scan over the usual parameter space, mentioned in section 5, and further required
that $m_{h_2}\leq$ 300 GeV, as corresponding production rates
become negligible for heavier masses. We used CalcHEP \cite{CalcHEP} to determine the cross sections for NMSSM $h_{1, 2}$ production for the following two processes:
\begin{equation}
gg\to b\bar b~{h_1} \phantom{aa} {\rm and} \phantom{aa}   gg\to b\bar b~{h_2},
\label{eq:proc}
\end{equation}
which were computed separately (i.e., without the interferences emerging whenever $h_1$ and $h_2$ have the same decay products).

The lightest two CP-even neutral Higgs boson masses are given by Eq.(25).
Recall though that the equation is at tree level, mainly for guidance in interpreting the upcoming figures, while NMSSMTools includes radiative corrections as well.

Figure \ref{fig:4m1,2-scan3} shows the correlations between all three Higgs masses
$m_{a_1}$, $m_{h_1}$ and $m_{h_2}$. Since the successful points emerging from the scan have small
values of $\lambda$, $\kappa$ and also $A_{\kappa}$, only rather small values of $m_{a_1}$ are allowed.
It is remarkable that the smaller $m_{a_1}$, the smaller $m_{h_1}$ and $m_{h_2}$ too (two top-panes). 
In the bottom-pane of the same figure, for $m_{h_2}$ around 120 GeV, $m_{h_1}$
can have values from just above 0 up to slightly less than 120 GeV, showing the possibility that the two Higgs
states can have the same mass, i.e., $m_{h_1} \sim m_{h_2}$. 
Notice also that the majority of points have $m_{h_1}$ between 115 and 120 GeV.

\begin{figure}
 \centering\begin{tabular}{cc}

 \includegraphics[scale=0.70]{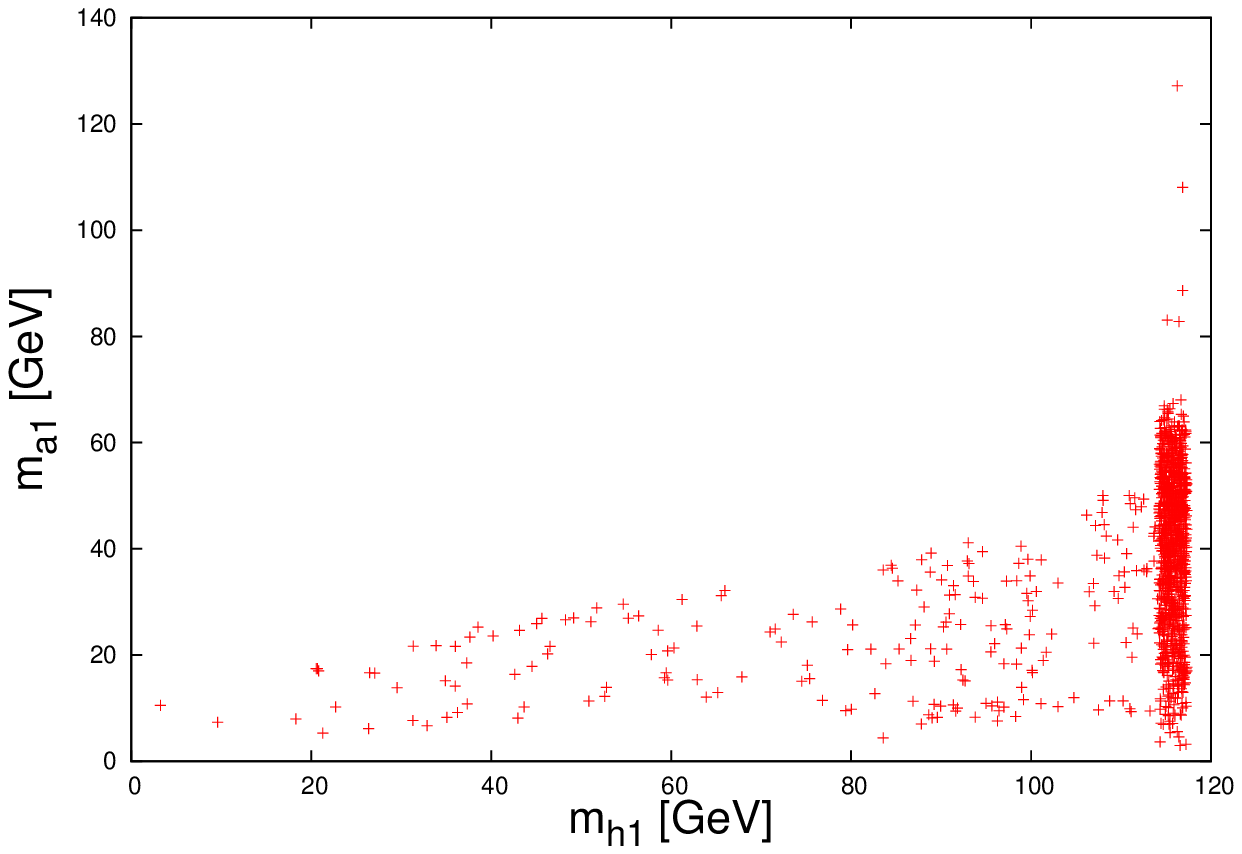}\\
\includegraphics[scale=0.70]{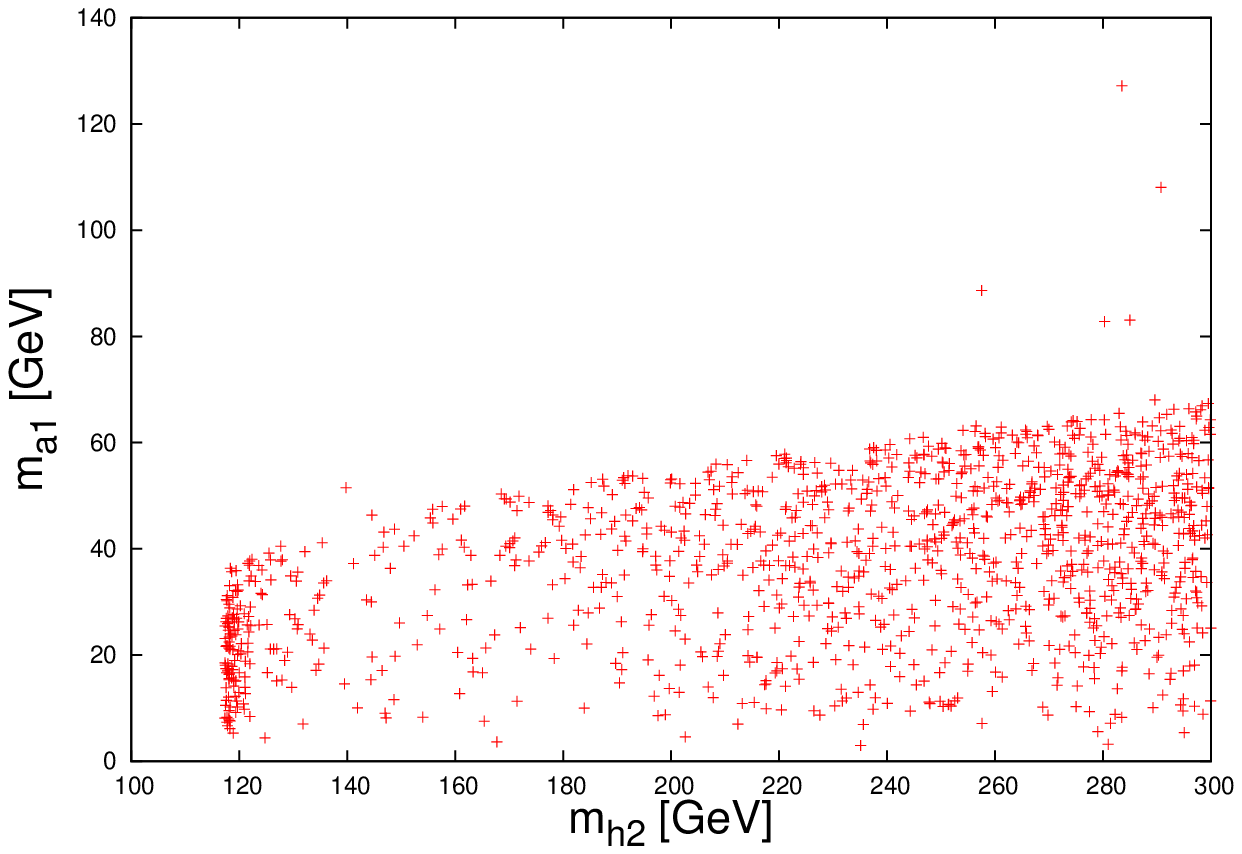}\\
  \includegraphics[scale=0.70]{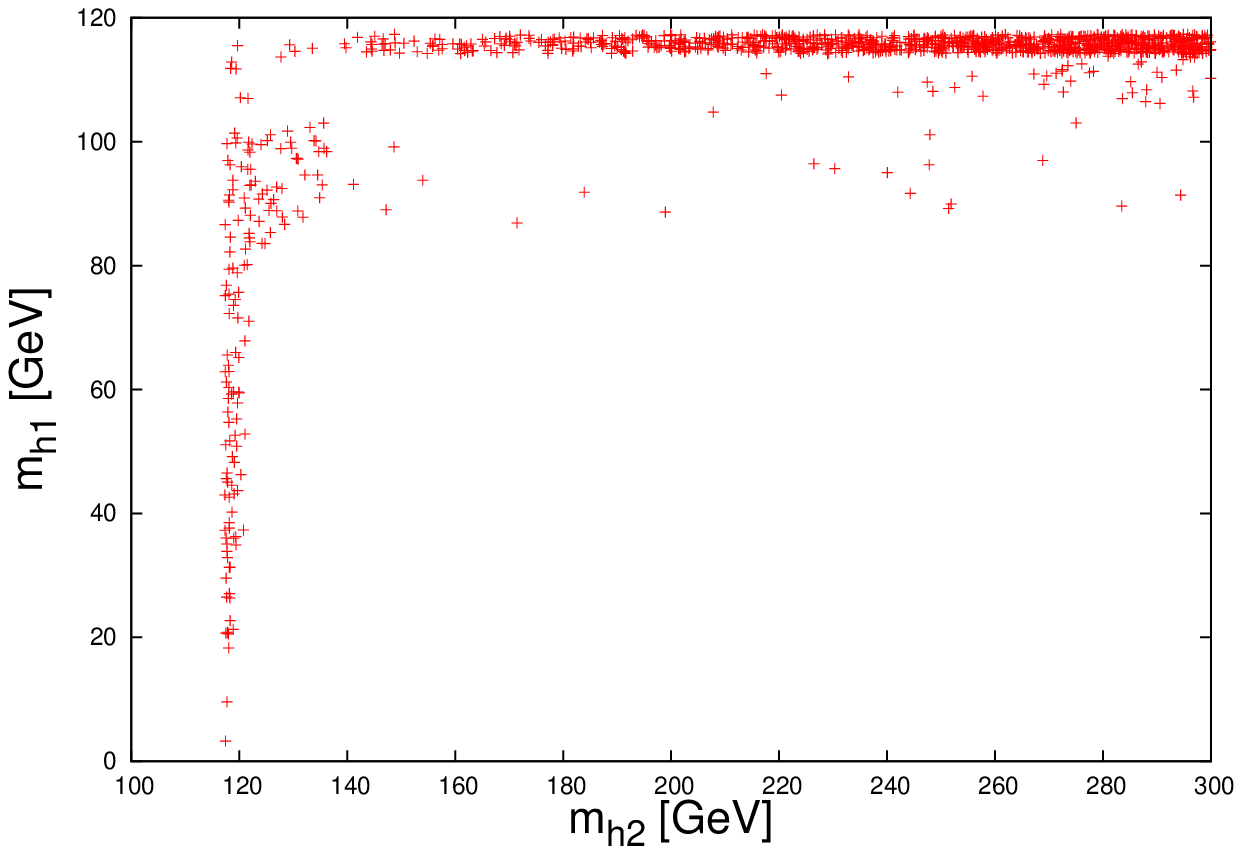}
 \end{tabular}
\caption{The correlations between the lightest CP-odd Higgs mass $m_{a_1}$
and the lightest two CP-even Higgs masses $m_{h_1}$ and $m_{h_2}$ and between the latter two.} 
\label{fig:4m1,2-scan3}   

\end{figure}

\subsection{Production of $h_1$ and $h_2$ decaying into two lighter Higgs bosons}

\begin{figure}
 \centering\begin{tabular}{cc}
  \includegraphics[scale=0.65]{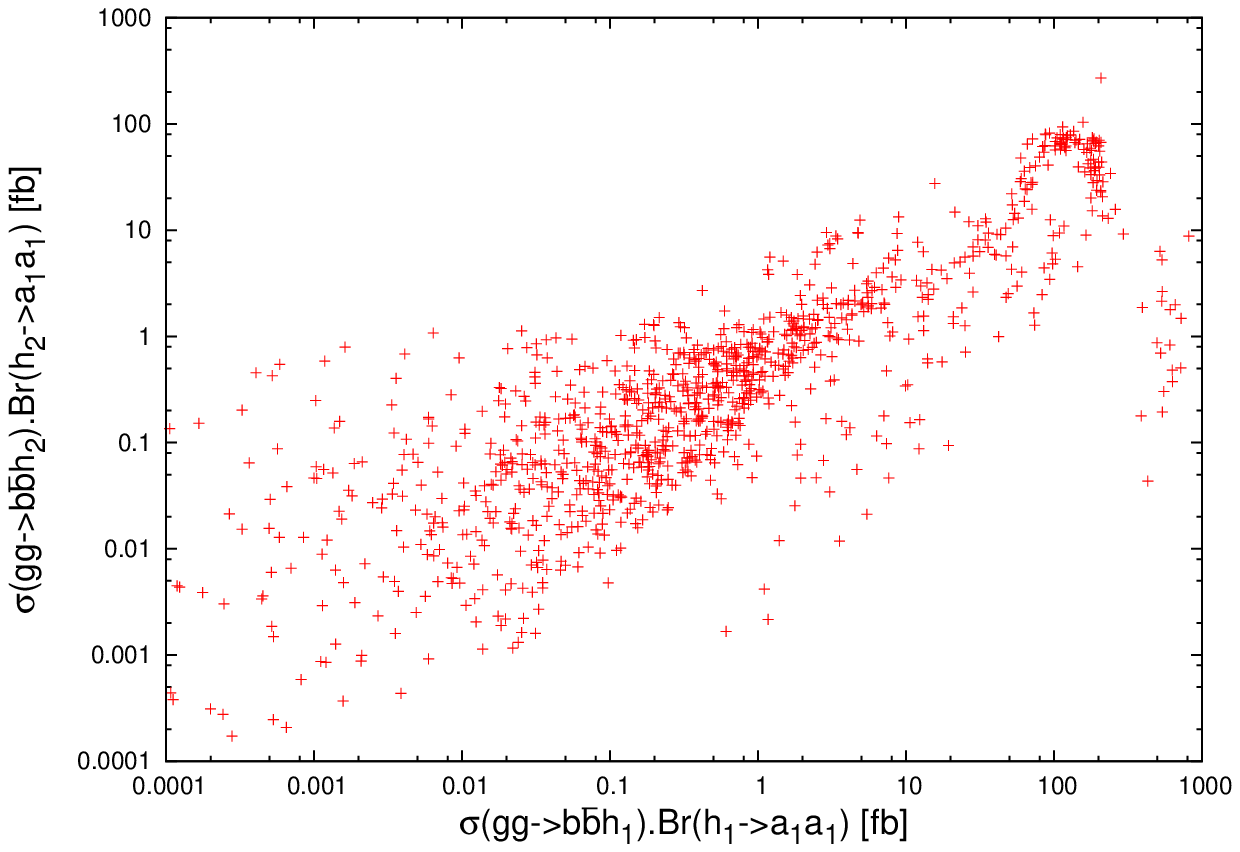}\\
   \includegraphics[scale=0.65]{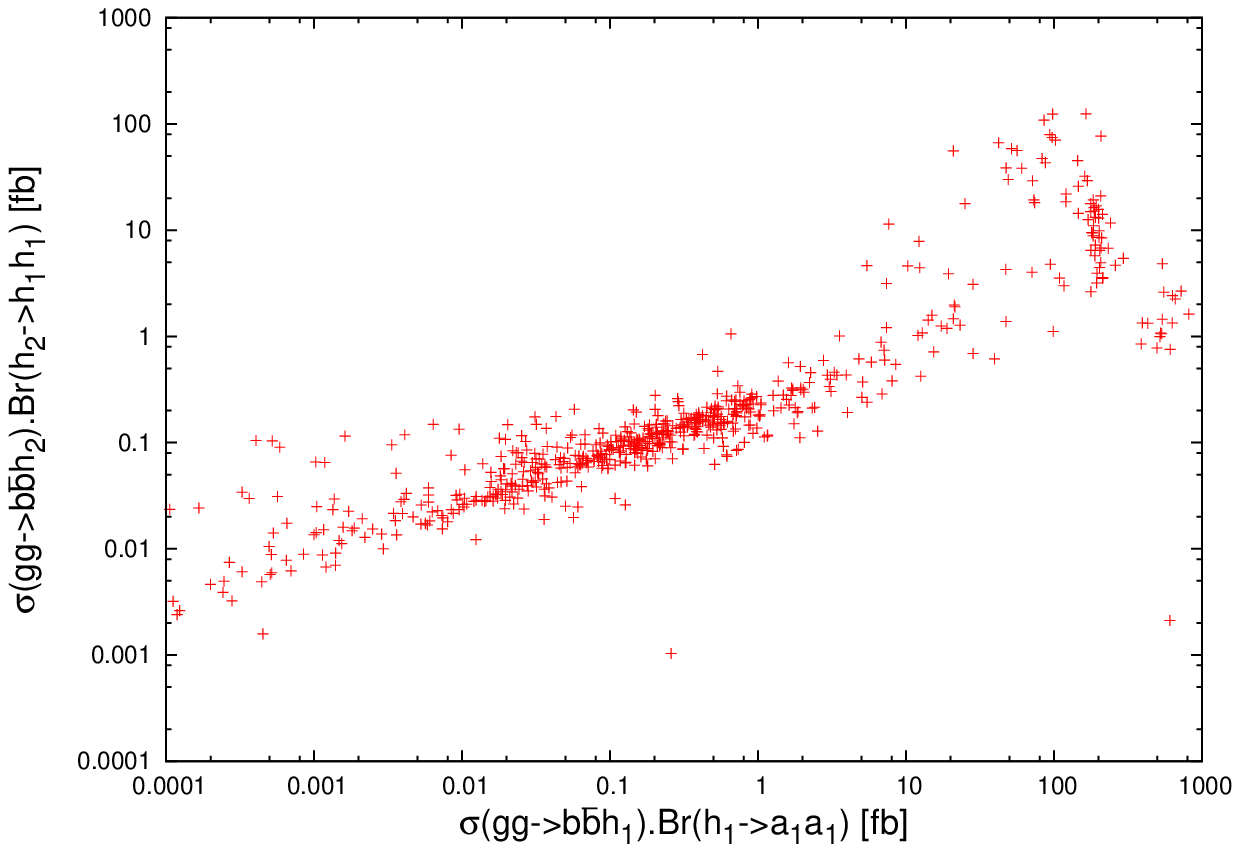}\\
  \includegraphics[scale=0.65]{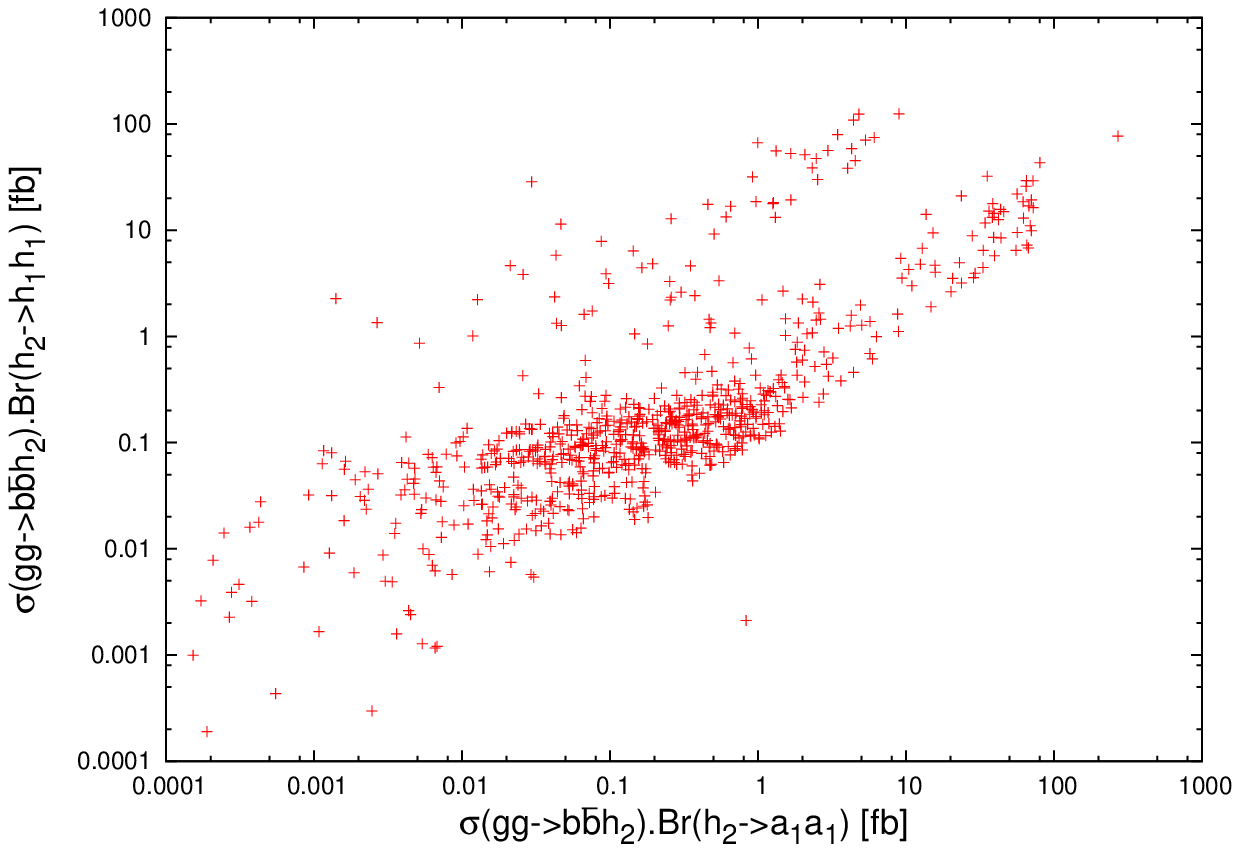}
     
 \end{tabular}
  \caption{The rates for 
$\sigma(gg\to b\bar b {h_1})~{\rm Br}(h_1\to a_1a_1)$  versus  
$\sigma(gg\to b\bar b {h_2})~{\rm Br}(h_2\to a_1a_1)$, 
$\sigma(gg\to b\bar b {h_1})~{\rm Br}(h_1\to a_1a_1)$  versus  
$\sigma(gg\to b\bar b {h_2})~{\rm Br}(h_2\to h_1h_1)$  and for
$\sigma(gg\to b\bar b {h_2})~{\rm Br}(h_2\to a_1a_1)$  versus  
$\sigma(gg\to b\bar b {h_2})~{\rm Br}(h_2\to h_1h_1)$.}
 \label{fig:4sigma-correlations5}

\end{figure}

The production times decay rates of $h_1$ and $h_2$, in which $h_1$ decays into two lighter $a_1$'s
and $h_2$ decays into either a pair of $a_1$'s or a pair of $h_1$'s, are shown in figure \ref{fig:4sigma-correlations5}.
This figure displays all the correlations between the three discussed production and decay processes.
 It is quite remarkable that the overall trend, despite an obvious spread also in the horizontal and
vertical directions, is such that when one channel grows in event yield there is also another one which also does,
 hence opening up the possibility of the simultaneous discovery of several Higgs states of the NMSSM 
(three neutral Higgses at the same time: $h_1$, $h_2$ and $a_1$), an exciting 
prospect in order to distinguish the NMSSM Higgs sector from the MSSM
one (in fact, a clear manifestation of a possible More-to-gain theorem being established){\footnote{A partonic signal-to-background
($S/B$) analysis for several $a_1a_1$ and $h_1h_1$ decays is currently being done in \cite{progress}.}}. 
   
\subsection{Production of $h_1$ and $h_2$ decaying into a gauge boson and a light CP-odd Higgs}
\begin{figure}
 \centering\begin{tabular}{cc}
 \includegraphics[scale=1]{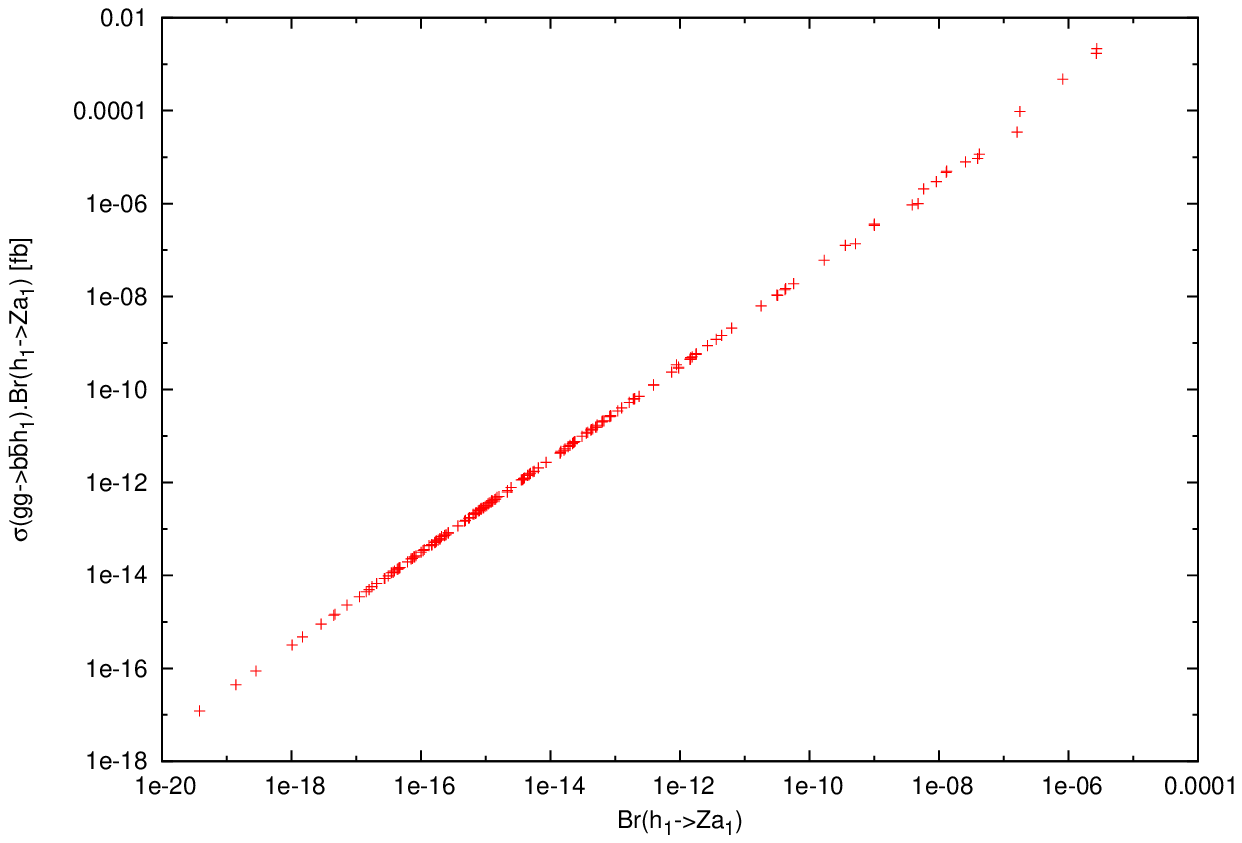}\\
\includegraphics[scale=1]{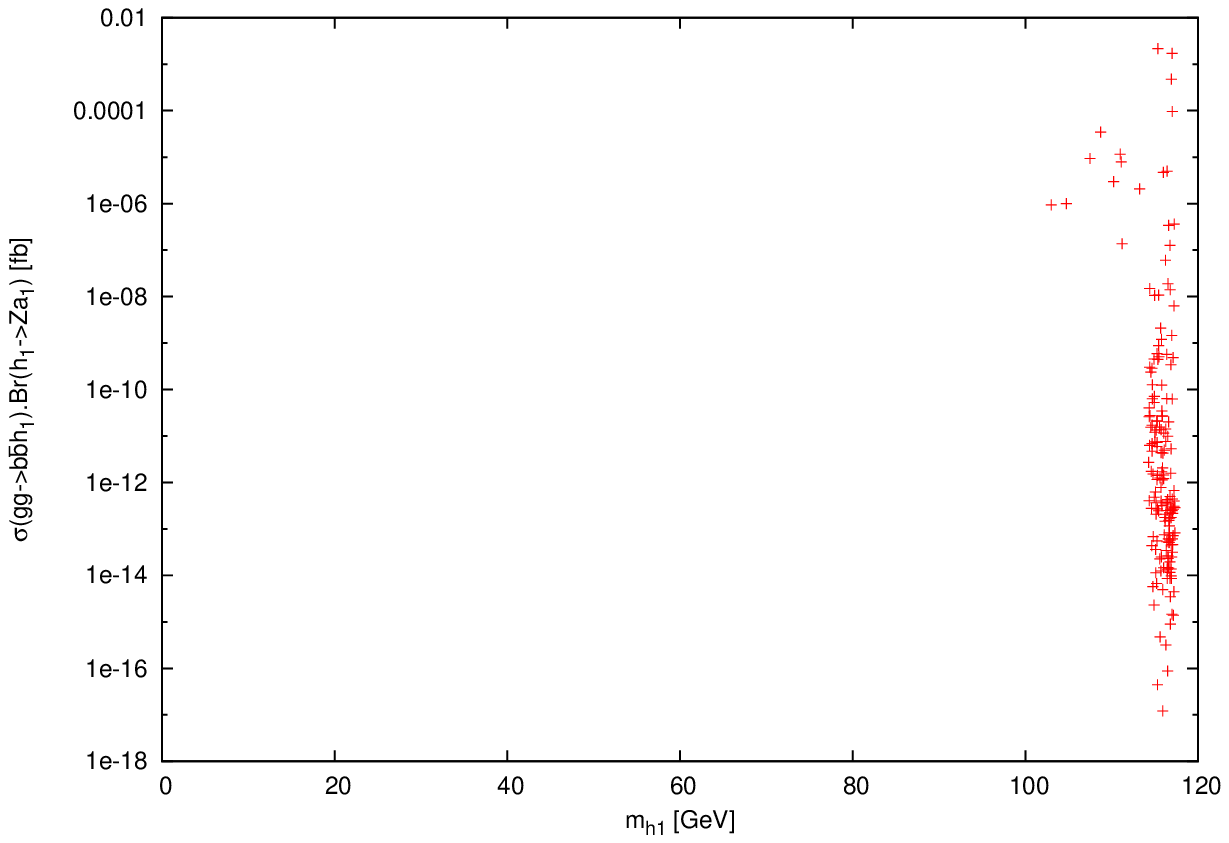}
 \end{tabular}
  \caption{The signal rate for $\sigma(gg\to b\bar b {h_1})~{\rm Br}(h_1\to Za_1)$ 
   as a function of the Br$(h_1\to Za_1)$ and of $m_{h_1}$.}
 \label{fig:4sigma-scan9}
\end{figure}

In this subsection, we examine the LHC discovery potential of the lightest two CP-even Higgs states $h_{1, 2}$,
followed by the decay $h_{1, 2}\to Za_1$.
Figure \ref{fig:4sigma-scan9} shows that the production rate
for the $h_1$, produced in association with a $b\bar b$ pair, is small, topping the 0.001 $fb$ level.
Such a production rate is presumably not enough to discover the $h_1$ at the LHC. The top-pane of the figure shows that there is 
a linear relation between the $h_1$ production rate and the Br($h_{1}\to Za_1$) because the production rate
$\sigma(gg\to b\bar bh_1)$ is nearly constant in our parameter space, which has large tan$\beta$. The bottom-pane of the figure shows that the points
passing the constraints have $m_{h_1}>100$ GeV.

\begin{figure}
 \centering\begin{tabular}{cc}
 \includegraphics[scale=0.7]{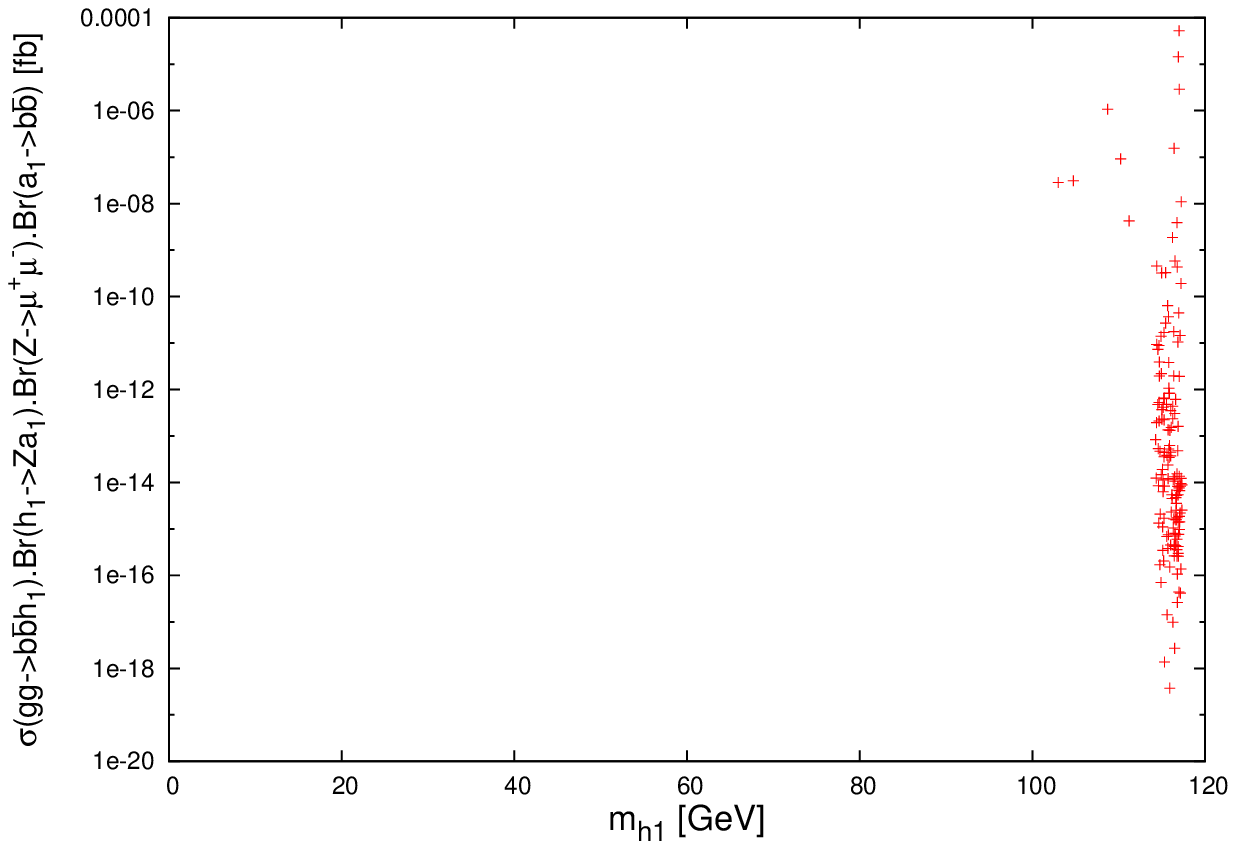}\\
\includegraphics[scale=0.7]{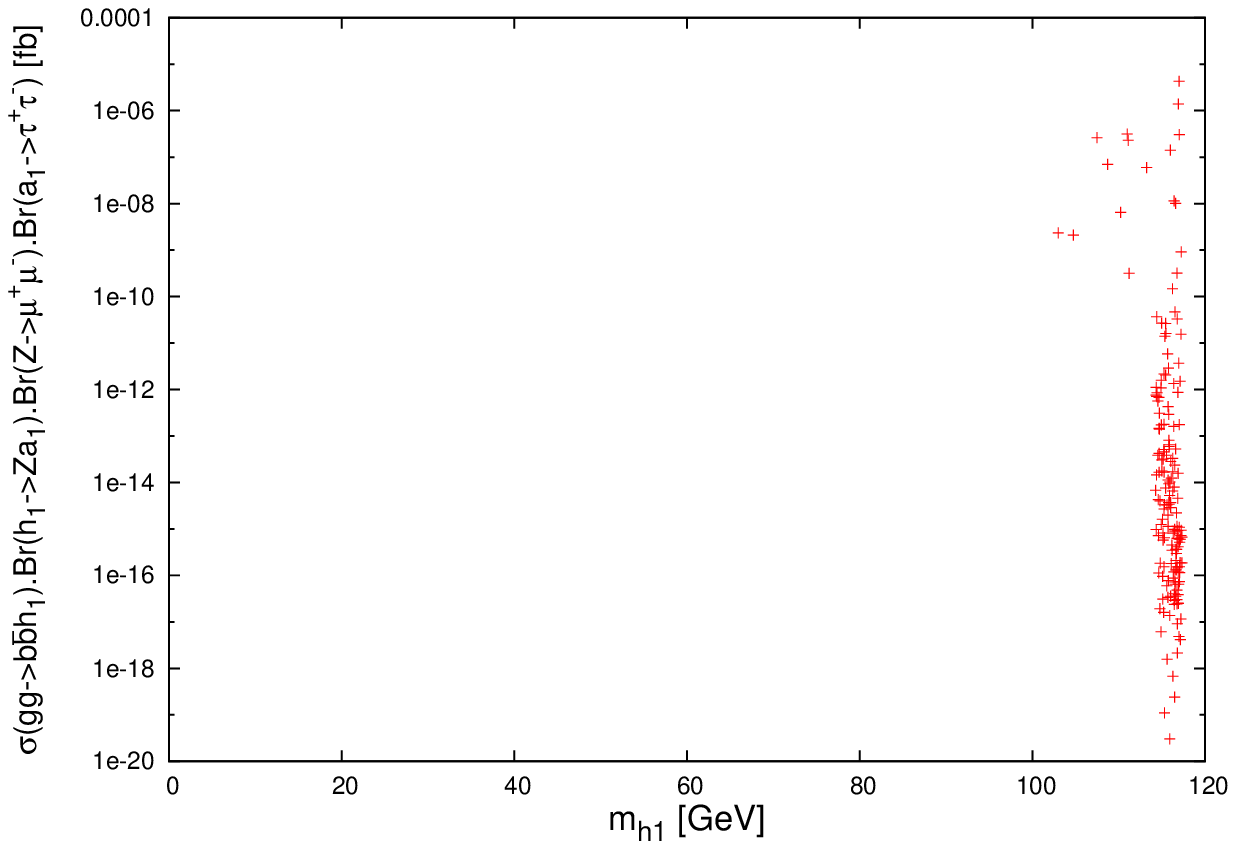}\\
\includegraphics[scale=0.7]{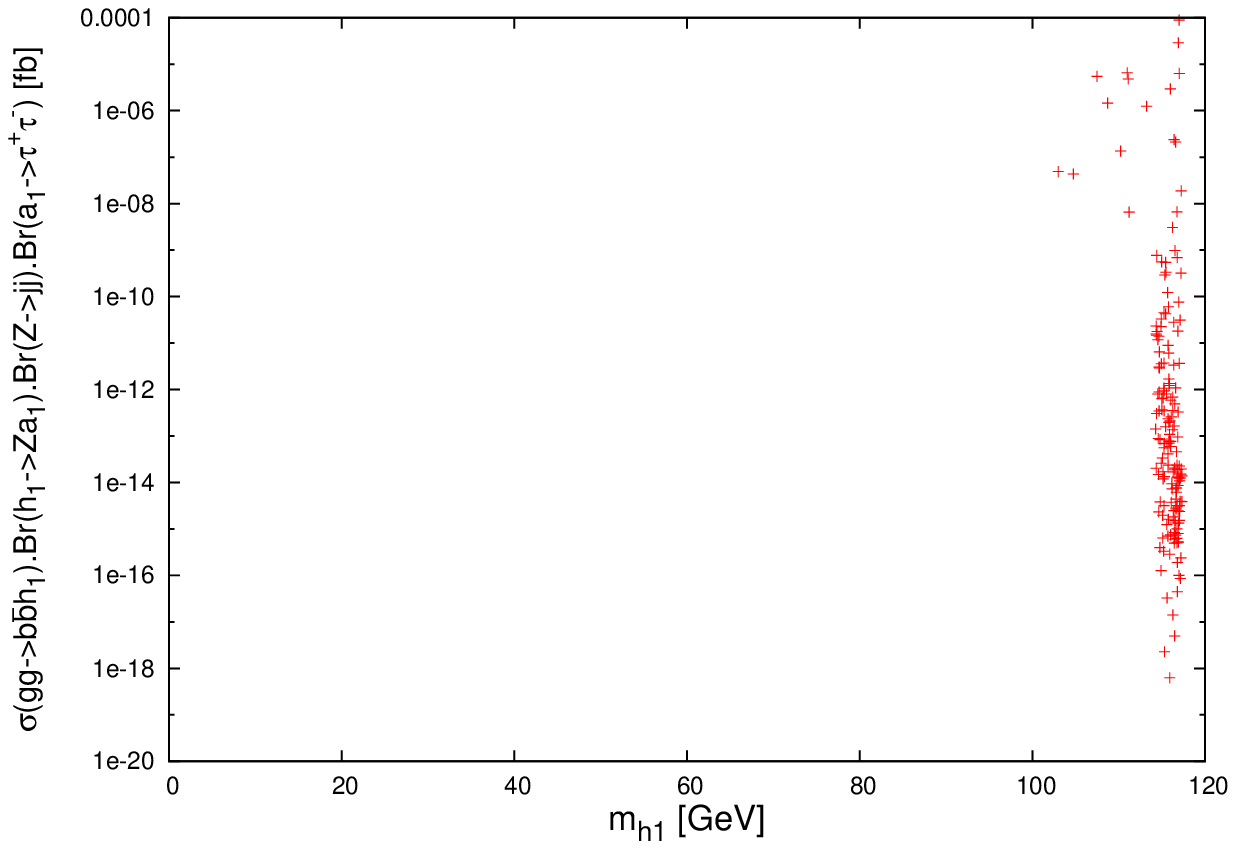}
       
 \end{tabular}
  \caption{The signal rate for $\sigma(gg\to b\bar b {h_1})~{\rm Br}(h_1\to Za_1)$ times ${\rm Br}(Za_1\to \mu^+\mu^-b\bar b)$,
 times ${\rm Br}(Za_1\to \mu^+\mu^-\tau^+\tau^-)$ and times ${\rm Br}(Za_1\to jj\tau^+\tau^-)$
  as functions of $m_{h_1}$.}
 \label{fig:4sigma-scan10}
\end{figure}

Figure \ref{fig:4sigma-scan10} illustrates the inclusive $h_1$ production rates ending up with $Za_1\to \mu^+\mu^-b\bar b$,
$Za_1\to \mu^+\mu^-\tau^+\tau^-$ and $Za_1\to jj\tau^+\tau^-$ (where $j =$ jet). It is clear that the production and decay 
rates are definitely too small, topping $10^{-4}$ $fb$ for the first and last channels and $10^{-5}$ $fb$ for the second one. 
Such rates are obviously not enough to discover the $h_1$ neither at the LHC nor at the SLHC with $1000$ $fb^{-1}$ of luminosity.
  
\begin{figure}
 \centering\begin{tabular}{cc}
 \includegraphics[scale=1]{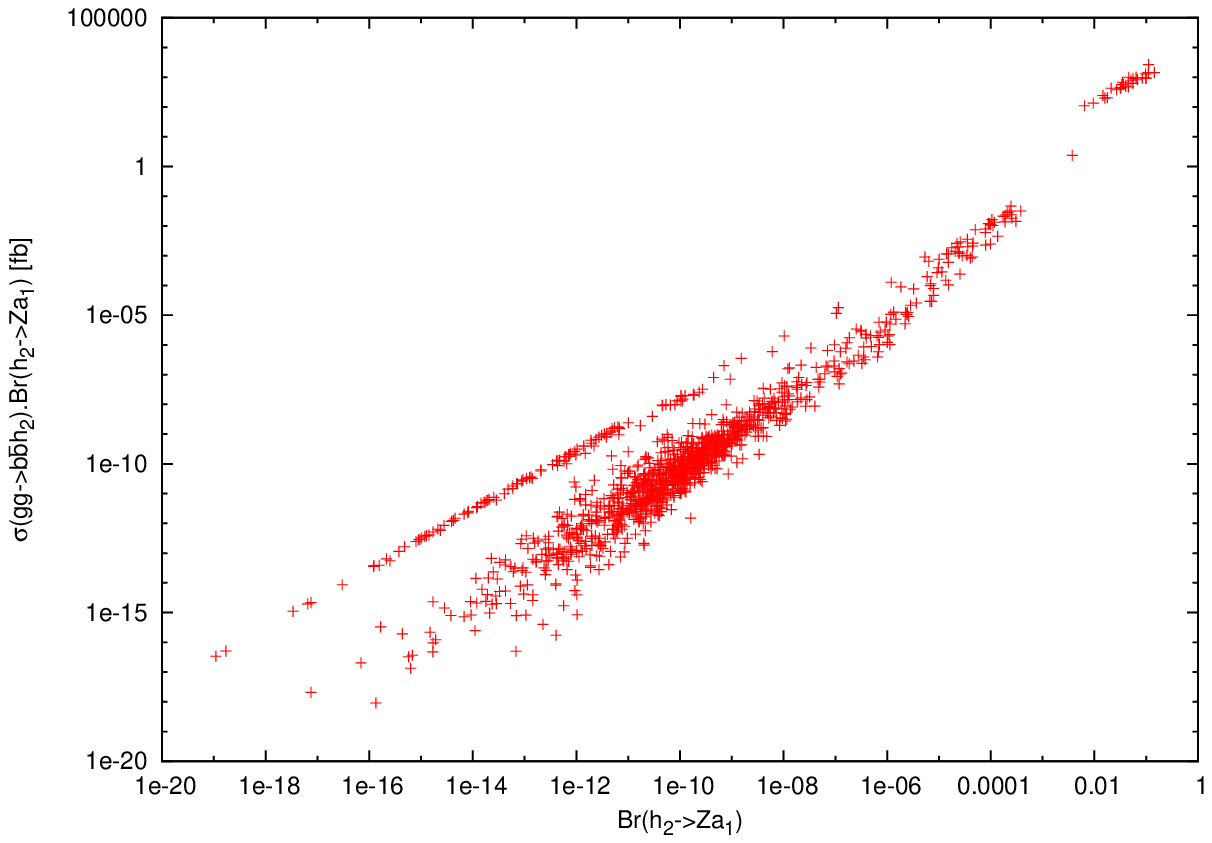}\\
\includegraphics[scale=1]{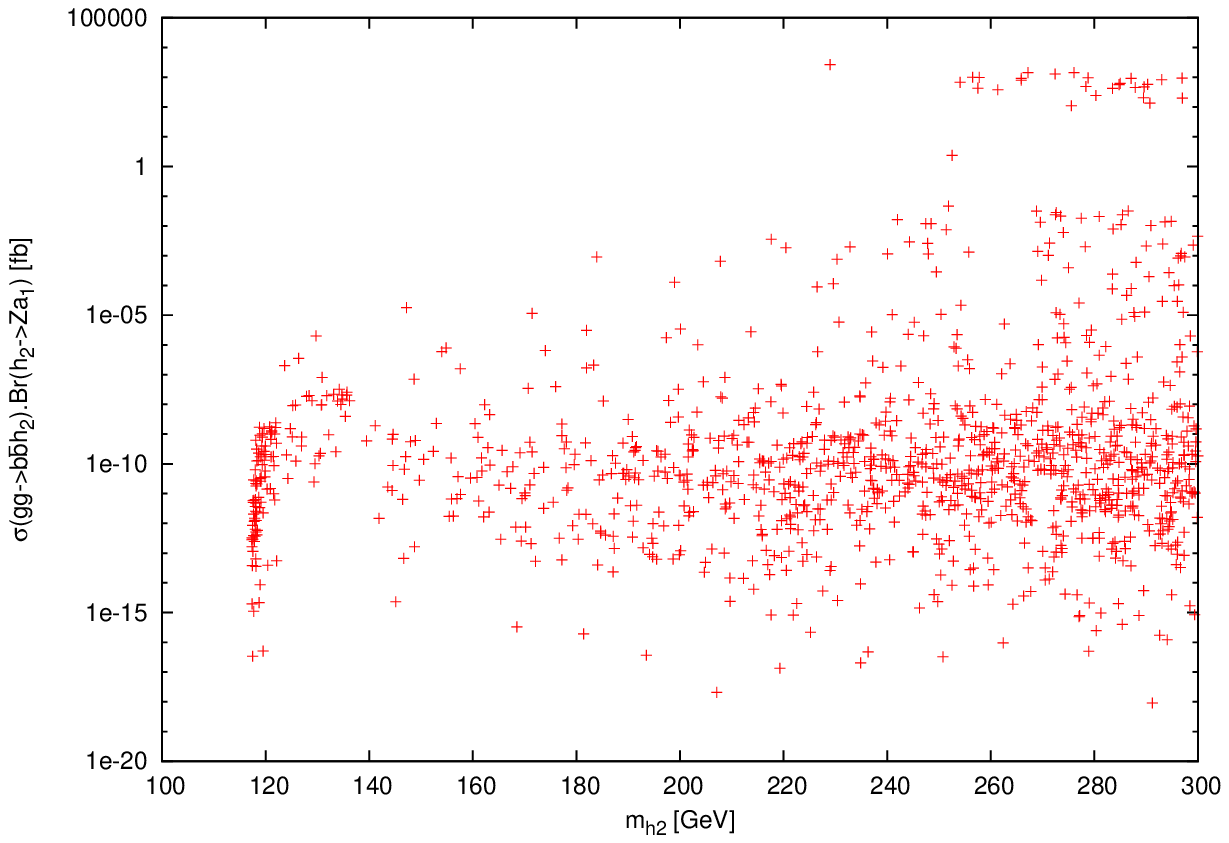}
       
 \end{tabular}
  \caption{The signal rate for $\sigma(gg\to b\bar b {h_2})~{\rm Br}(h_2\to Za_1)$
   as a function of the Br$(h_2\to Za_1)$ and of $m_{h_2}$.}
 \label{fig:4sigma-scan11}
\end{figure}

In contrast, the situation for $h_2$ is promising as one can notice that $\sigma(gg\to b\bar bh_2){\rm Br}(h_2\to Za_1)$
is sizable, topping the 10000 $fb$ level (figure \ref{fig:4sigma-scan11}). The highest values of the cross section are accompanied by 
an intriguingly large Br($h_2\to Za_1$), reaching up to $10$\%. It is clear from the top-pane of the figure
that the distribution over the branching ratio for $h_2$ is not as uniform as that for the $h_1$ because
the production rate $\sigma(gg\to b\bar bh_2)$ depends strongly on the tree level parameters unlike that for
$h_1$. The bottom-pane of the figure shows that the highest cross section occurs for
$m_{h_2}>220$ GeV. 

\begin{figure}
 \centering\begin{tabular}{cc}
 \includegraphics[scale=0.7]{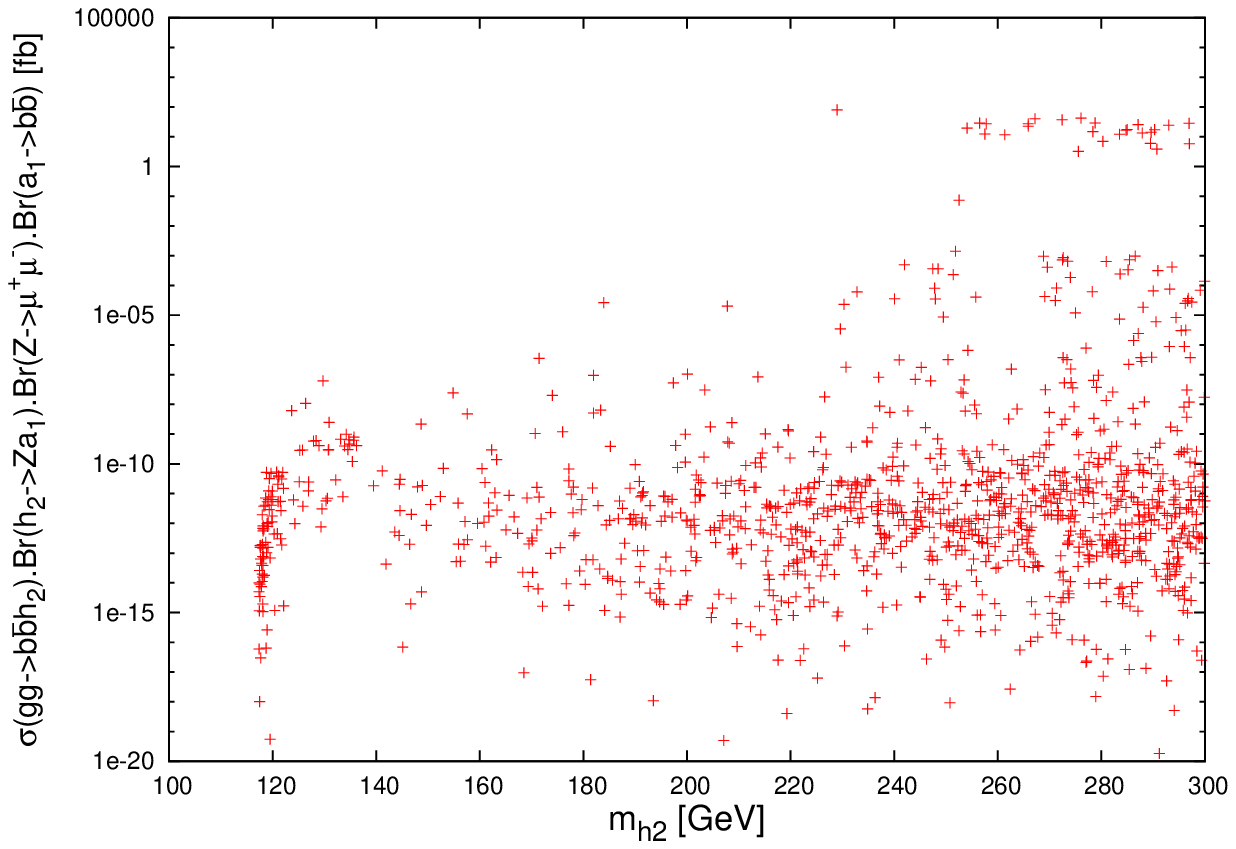}\\
\includegraphics[scale=0.7]{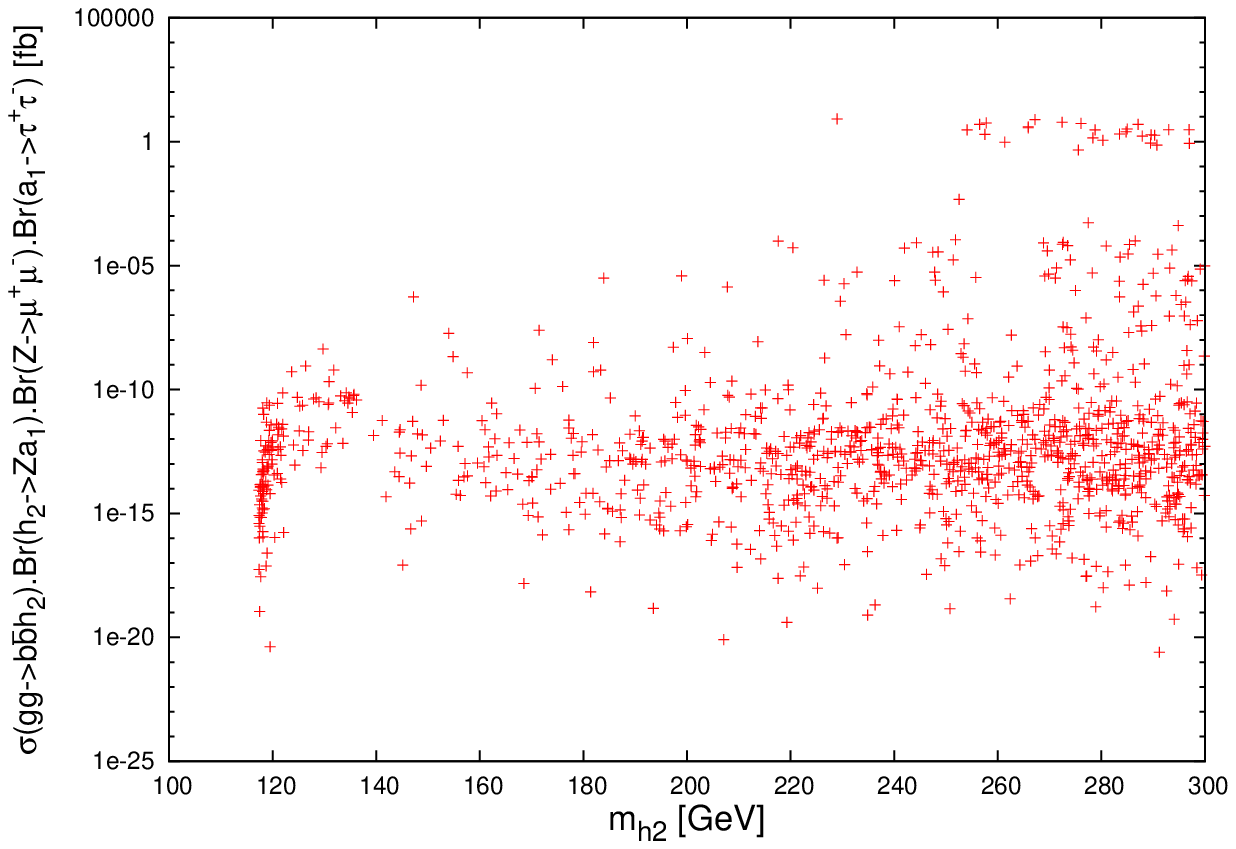}\\
\includegraphics[scale=0.7]{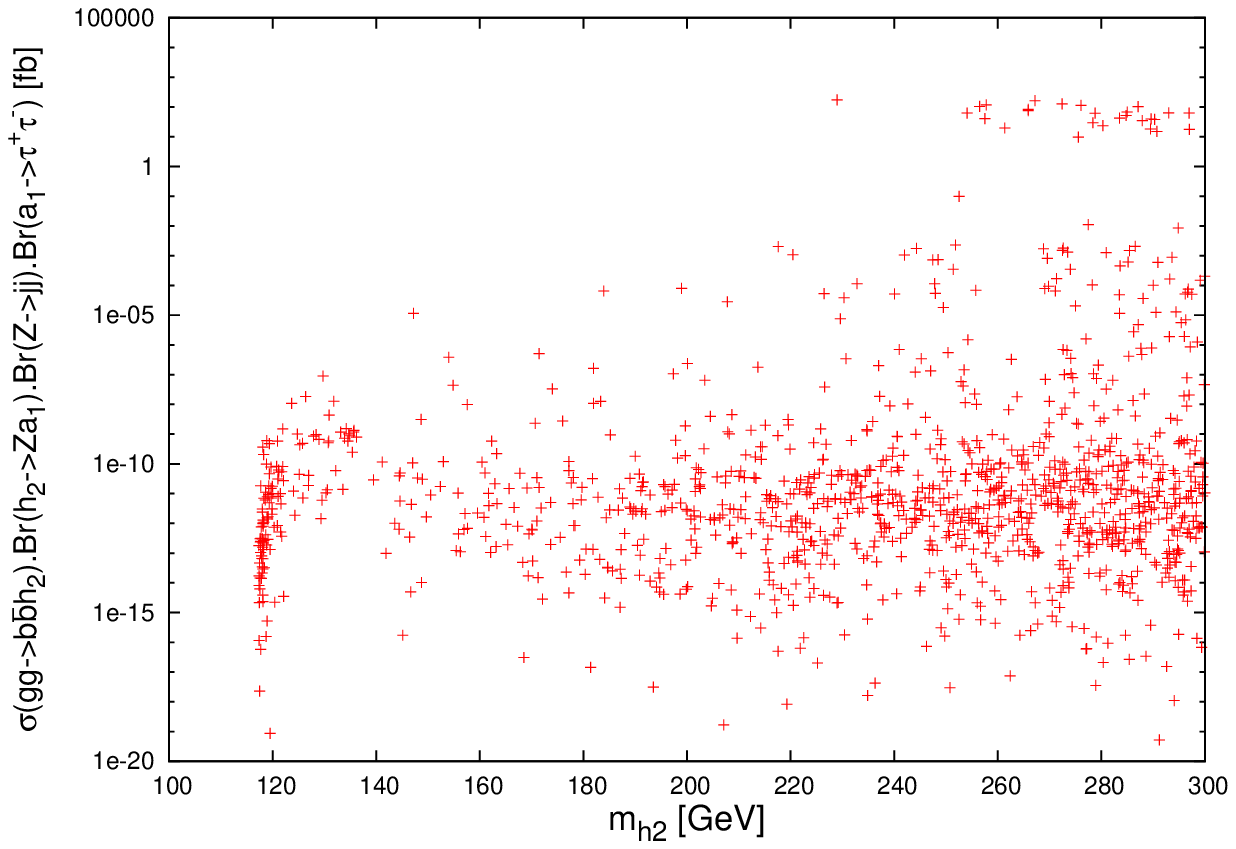}
       
 \end{tabular}
  \caption{The signal rate for $\sigma(gg\to b\bar b {h_2})~{\rm Br}(h_2\to Za_1)$ times ${\rm Br}(Za_1\to \mu^+\mu^-b\bar b)$,
 times ${\rm Br}(Za_1\to \mu^+\mu^-\tau^+\tau^-)$ and times ${\rm Br}(Za_1\to jj\tau^+\tau^-)$
 as functions of $m_{h_2}$.}
 \label{fig:4sigma-scan12}
\end{figure}

In order to study the detectability of $h_2$ decaying into a gauge boson and a light CP-odd Higgs state at the LHC, we have calculated the inclusive
production rates ending up with $\mu^+\mu^-b\bar b$, $\mu^+\mu^-\tau^+\tau^-$ and $jj\tau^+\tau^-$ (figure \ref{fig:4sigma-scan12}).
The event rates for these processes are
at the $\cal {O}$(100) $fb$ level at the most. While clearly this number
is not very large, signal events may still be detectable at
planned LHC luminosities, especially if the background
can be successfully reduced to manageable levels{\footnote{A partonic signal-to-background
($S/B$) analysis for $jj\tau^+\tau^-$ final state has been done in \cite{Almarashi:2011qq}, showing very promising results.}}. 
In short, there is a small
but well defined region of the NMSSM parameter space where the $h_2$
and $a_1$ states, both with a mixed singlet
and doublet nature, could potentially be detected at
the LHC if 220 GeV $\lesssim$ $m_{h2}$ $\lesssim$ 300 GeV and 15 GeV
$\lesssim$ $m_{a1}$ $\lesssim$ 60 GeV, in the $h_2\to Za_1\to \mu^+\mu^-b\bar b$, $h_2\to Za_1\to \mu^+\mu^-\tau^+\tau^-$ and
 $h_2\to Za_1\to jj\tau^+\tau^-$  modes,
when the CP-even Higgs state is produced in association
with a $b\bar b$ pair for rather large tan$\beta$.

\section{\bf Conclusions}
The NMSSM has a singlet Superfield in addition to
the usual Higgs doublets of the MSSM. This singlet gives rise to a more varied phenomenology in the case of the NMSSM,
compared to that of the MSSM. For instance, this singlet Superfield mixes with the neutral components
of the doublets, giving rise to one CP-even Higgs, one CP-odd Higgs and one extra neutralino in
addition to the usual spectrum of the MSSM.
Therefore,
in the NMSSM, by assuming CP-conservation, there are seven Higgses: three CP-even, two CP-odd and
a pair of charged Higgses. We have investigated whether or not at least one Higgs boson of the NMSSM can
be discovered at the LHC (`No-lose theorem') and/or is possible to find some regions in 
the parameter space where more and/or different Higgs states of the NMSSM are detectable at the
LHC, compared to those available within the MSSM (`More-to-gain theorem').

Because of the mixing between the Higgs singlet and doublets, Higgs-to-Higgs decays are kinematically
possible for large regions of the NMSSM parameter space even for small masses of the Higgs states,
which is impossible in the MSSM. For instance, a SM-like Higgs can decay into a pair of the lightest NMSSM CP-odd
Higgses. This decay can be dominant in sizable areas of the NMSSM parameter space. Such a decay
has a significant meaning if one notices that it can explain a $2.3 \sigma$ excess occurred
at LEP for the process $e^+e^-\to Zb\bar b$ for $M_{b\bar b}$$\sim 98$ GeV and the $2.6\sigma$ excess recently emerged at the LHC (primarily in the $\gamma\gamma$ decay mode). Moreover, a SM-like
Higgs with mass of order 100 GeV, which has no-fine tunning, can naturally occur in the NMSSM
and this scenario is preferred by precision EW data. In addition, the NMSSM can solve both
the $\mu$-problem and the little hierarchy problem of the MSSM.

In the context of the NMSSM, we have proven that a very light CP-odd Higgs state with mass $m_{a_1}\lesssim M_Z$, which has a large
singlet component and a small doublet one, can be discovered at the LHC via
Higgs production in association with a bottom-antibottom pair. This mode is dominant at large tan$\beta$. After performing
several analyses for signals and dominant backgrounds, not documented here yet referred to, we have proven that this production mode is the ideal one to discover the $a_1$ through the following signatures: $(i)$ $\tau^+\tau^-$ decay mode, in which
$a_1$ can be discovered with mass up to $M_Z$; $(ii)$ $\mu^+\mu^-$ decay mode, if $10\lesssim m_{a_1}\lesssim 60$ GeV.
Further, despite the fact that the $b\bar b$ decay mode is dominant in most regions of parameter space that have light $a_1$,
this channel has huge QCD background and a smaller signal-to-background ratio.
Finally, we also looked at the detectability of $a_1$ through the $\gamma\gamma$ decay mode but this
proved unuseful despite the fact that this decay mode can be dominant in some areas of the NMSSM parameter space. 

We believe that the results presented in sections 7 and 8  have a twofold relevance. Firstly, they support
the `No-lose theorem' by looking for direct
$a_1$ production rather than looking for its production through the decays $h_{1, 2}\to a_1a_1$, which may not give
a sufficient signal significance. Secondly, they corroborate the `More-to-gain theorem' as such very light $a_1$'s (with $m_{a_1}\lesssim M_Z$)
are not at all possible in the MSSM. Altogether, the existence of such a light neutral Higgs state is a direct evidence for the non-minimal
nature of the SUSY Higgs sector.

Finally, we have mentioned in section 9 the importance of Higgs-to-Higgs decays in the NMSSM, here occurring after Higgs boson production in association with $b\bar b$ pairs (unlike in most 
previous literature), and have shown that such
decays should be taken seriously before proving, or otherwise, the `No-lose theorem'. In fact, we also have shown that
such decays are dominant in sizable regions of the NMSSM parameter space. We have studied the LHC discovery potential of a CP-even
Higgs boson $h_1$ or $h_2$, decaying into a pair of light CP-odd Higgses $a_1$'s, and also $h_2$ decaying into a pair of $h_1$'s. We have found that these channels can give sizable
signal rates, which could allow one to detect simultaneously two Higgs bosons: $h_1$ and $a_1$, $h_2$ and $a_1$ or
$h_2$ and $h_1$. In addition, we have shown that the LHC has the potential to discover the three neutral Higgs bosons at the same time. Furthermore,
we have studied the LHC discovery potential for $h_1$ and $h_2$ decaying into $Za_1$ and have shown that, while the discovery of the $h_1$
through this channel is impossible, there is a small but well
defined region of the NMSSM parameter space where the $h_2$ state could potentially be discovered.

\section*{Acknowledgments}
M.M.A. gratefully acknowledges financial 
support from Taibah University in Saudi Arabia.
S.M. is financially partially supported through the NExT Institute.


\end{document}